\documentclass[fleqn,usenatbib]{mnras}

\usepackage[T1]{fontenc}
\usepackage{ae,aecompl}

\usepackage{graphicx}	
\usepackage{amsmath}	
\usepackage{amssymb}	

\usepackage{newtxtext,newtxmath}

\title[Self-lensing binary]{A self-lensing binary massive black hole interpretation of quasi-periodic eruptions}

\author[Adam Ingram et al]{
Adam Ingram,$^{1}$\thanks{E-mail: adam.ingram@physics.ox.ac.uk}
Sara E. Motta,$^{1}$
Suzanne Aigrain$^{1}$
and Aris Karastergiou$^{1}$
\\
$^{1}$Department of Physics, Astrophysics, University of Oxford, Denys Wilkinson Building, Keble Road, Oxford OX1 3RH, UK
}

\date{Accepted 2021 February 26. Received 2021 February 23; in original form 2020 November 16}

\pubyear{2021}

\begin{document}
\label{firstpage}
\pagerange{\pageref{firstpage}--\pageref{lastpage}}
\maketitle

\begin{abstract}
Binary supermassive black hole (SMBH) systems result from galaxy mergers, and will eventually coalesce due to gravitational wave (GW) emission if the binary separation can be reduced to $\lesssim 0.1$ pc by other mechanisms. Here, we explore a gravitational self-lensing binary SMBH model for the sharp (duration $\sim 1$ hr), quasi-regular X-ray flares -- dubbed quasi-periodic eruptions -- recently observed from two low mass active galactic nuclei: GSN 069 and RX J1301.9+2747. In our model, the binary is observed $\sim$edge-on, such that each SMBH gravitationally lenses light from the accretion disc surrounding the other SMBH twice per orbital period. The model can reproduce the flare spacings if the current eccentricity of RX J1301.9+2747 is $\epsilon_0 \gtrsim 0.16$, implying a merger within $\sim 1000$ yrs. However, we cannot reproduce the observed flare profiles with our current calculations. Model flares with the correct amplitude are $\sim 2/5$ the observed duration, and model flares with the correct duration are $\sim 2/5$ the observed amplitude. Our modelling yields three distinct behaviours of self-lensing binary systems that can be searched for in current and future X-ray and optical time-domain surveys: i) periodic lensing flares, ii) partial eclipses (caused by occultation of the background mini-disc by the foreground mini-disc), and iii) partial eclipses with a very sharp in-eclipse lensing flare. Discovery of such features would constitute very strong evidence for the presence of a supermassive binary, and monitoring of the flare spacings will provide a measurement of periastron precession.

\end{abstract}




\section{Introduction}

Remarkably sharp, high amplitude, regular flares -- dubbed quasi-periodic eruptions (QPEs) -- have recently been discovered in the X-ray light curves of two low mass active galactic nuclei (AGN): GSN 069 \citep{Miniutti2019} and RX J1301.9+2747 (hereafter RX J1301; \citealt{Giustini2020}). Both AGN have been optically classified as Seyfert 2 galaxies \citep{Sun2013,Shu2018}, and both exhibit a thermal X-ray spectrum consistent with a standard thin accretion disc \citep{Shakura1973} around a black hole (BH) of mass $M \sim 10^{5-6}~M_\odot$. In GSN 069, QPEs are observed in \textit{XMM-Newton} and \textit{Chandra} data from 2018 and 2019 to have a duration of $\sim 1$ hr and a spacing of $\sim 9$ hrs, and the peak $0.2-2$ keV count rate is $\sim 1$ order of magnitude higher than the quiescent level. There is evidence that the true period in GSN 069 is $\sim 18$ hours, with a larger flare being followed by a smaller flare just over 9 hrs later, followed by another flare just under 9 hrs later, and so on (Miniutti et al in prep). 

Similar flares in the X-ray light curve of RX J1301 had previously been seen in an \textit{XMM-Newton} observation from 2000, a \textit{Chandra} observation from 2009 and possibly \textit{ROSAT} observations \citep{Dewangan2000} in the early-mid 1990s (Earth occultations make confirmation challenging). However, since no single observation contained more than one full flare, it was not possible to determine whether or not the flares were periodic until a recent \textit{XMM-Newton} observation that revealed the same large-flare-followed-by-small-flare behaviour exhibited by GSN 069, except the long and short waiting times are in this case $\sim 5.6$ hrs and $\sim 3.6$ hrs respectively \citep{Giustini2020}. Since the full flare in the 2000 \textit{XMM-Newton} observation is preceded at the start of the exposure by the tail end of the previous flare $\sim 5$ hrs earlier, it appears that the QPE spacings in RX J1301 have evolved over a 19 year timescale.

Such sharp flares are unlikely to be driven by changes in the accretion rate onto the BH, since even a $\delta$-function perturbation in accretion rate will be smoothed on approximately the viscous timescale by the effective viscosity of the disc \citep{Lynden-Bell1974,Frank2002,Mushtukov2018}. The viscous timescale for a disc around an $M \sim 10^5~M_\odot$ BH is hundreds of hrs, which is far larger than the $\sim 1$ hr duration of the QPE flares. Although variability on timescales shorter than the viscous timescale is possible, short accretion flares are expected to have a fast rise with an exponential decay, whereas the QPE flares are remarkably symmetric \citep{Miniutti2019}.

Here we explore a model whereby GSN 069 and RX J1301 are both close binary massive BH systems. Such systems are thought to be surrounded by a circumbinary disc with inner edge $\approx 2 r_a(1+\epsilon)$ \citep[e.g.][]{Pringle1991,MacFadyen2008,D'Orazio2013}, where $r_a$ and $\epsilon$ are respectively the semi-major axis and eccentricity of the binary, and each BH is thought to be surrounded by its own mini-disc \citep[e.g.][]{Farris2014,Shi2015,Ryan2017,Bowen2017,Bowen2018,d'Ascoli2018}. If the system is viewed from a high inclination angle (i.e. nearly edge-on), one BH will pass in front of the other twice per orbit, leading the foreground BH to lens light from the background BH's mini-disc \citep{D'Orazio2018}. There will therefore be two flares per orbit. The implied orbital periods ($P \sim 18$ hrs for GSN 069 and $P \sim 9$ hrs for RX J1301) are very short, requiring both systems considered here to be very close binaries ($r_a$ {as small as} hundreds of gravitational radii). The larger flare will occur when the primary acts as the lens for two reasons: 1) the more massive BH is a stronger gravitational lens, and 2) the accretion rate onto the secondary is thought to be larger than that onto the primary \citep{Farris2014,Duffell2020}. The latter point also implies that the mass ratio, $q = M_2/M_1$, is likely to have been driven close to unity by enhanced accretion onto the secondary over the long period of time taken for gravitational wave (GW) emission to evolve these systems to such a small separations. This provides a natural explanation for why the amplitude of the large flare is only just greater than that of the small flare in both objects. The long and short waiting times between flares can be explained by an eccentric orbit (although a small eccentricity is expected after prolonged accretion; \citealt{Zrake2020}). {Our model is similar to one recently applied to the AGN nicknamed `Spikey' \citep{Hu2020,Kun2020}, except in that case the inferred orbital period was far longer ($\sim 418$ days).} In this paper, we explore two tests of our binary interpretation of QPEs. In Section \ref{sec:evolution} we study the evolution of the QPE flares in RX J1301 over the 19 years from 2000 to 2019, and in Section \ref{sec:profiles} we develop a quantitative model for the self-lensing flare profiles and compare the properties of the model to those of the GSN 069 QPE flares. We discuss our results in Section \ref{sec:discussion} and finally conclude in Section \ref{sec:conclusions}.

\begin{figure}
\centering
\includegraphics[width=8cm,trim=11.0cm 6.0cm 4.5cm 7.0cm,clip=true]{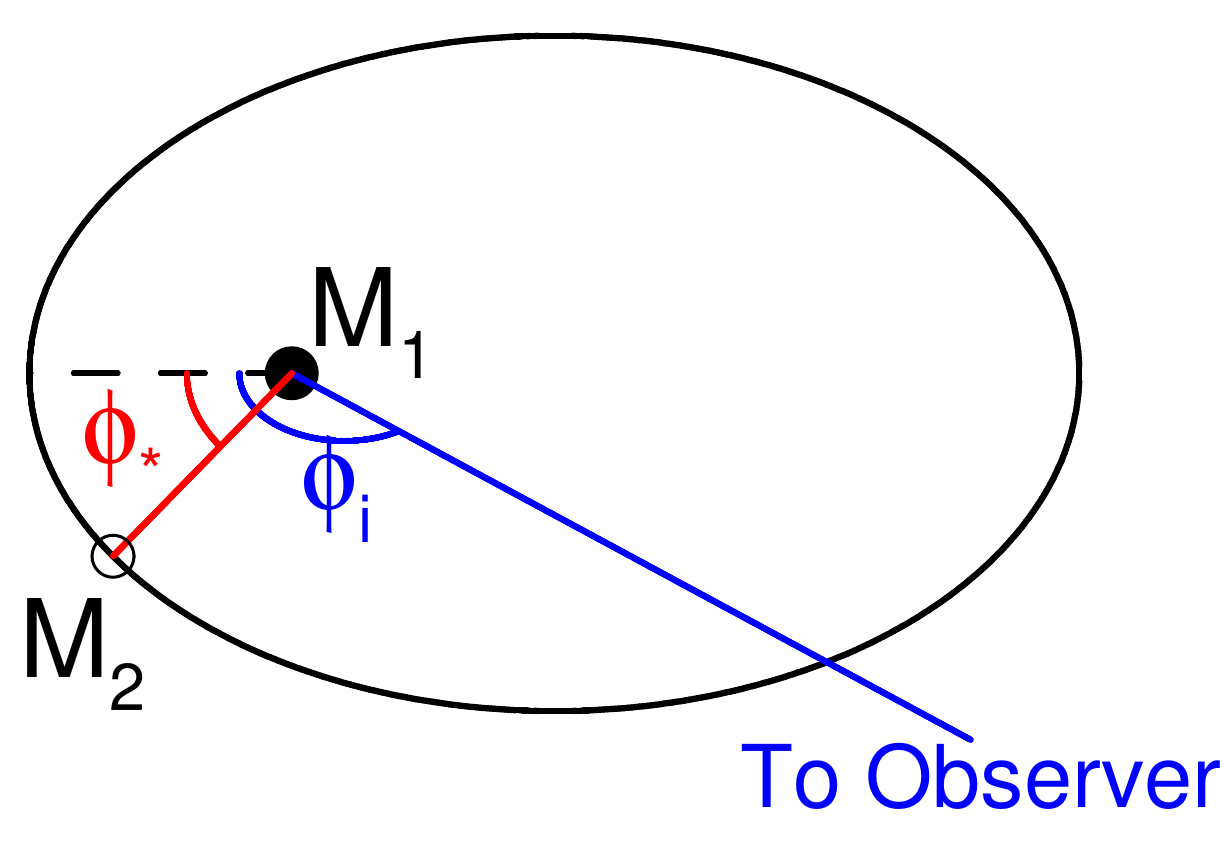}
\caption{Binary system viewed from above the orbital plane in the rest frame of the primary, such that the secondary (open circle) orbits the primary (filled circle) with orbital phase $\phi_*$ defined from peri-center (dashed line). The viewer azimuth $\phi_i$ is the angle between peri-centre and the projection of the observer's line of sight on the orbital plane (blue line).}
\label{fig:ellipse}
\end{figure}

\begin{figure}
\centering
\includegraphics[width=8cm,trim=1.0cm 1.5cm 2.0cm 2.5cm,clip=true]{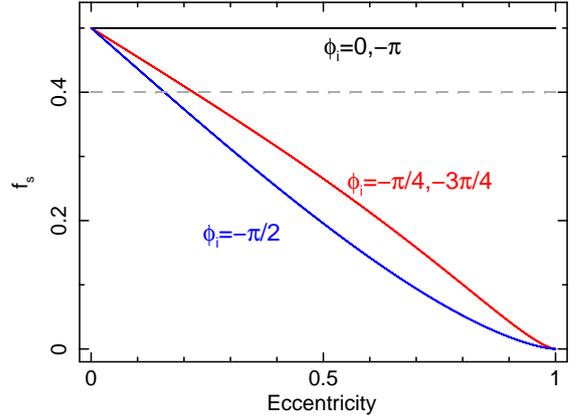}
\caption{Fraction of the orbit occupied by the waiting time from the small flare (secondary is the lens) to the large flare (primary is the lens) as a function of orbital eccentricity, for different values of observer azimuth, $\phi_i$ (defined in Fig. \ref{fig:ellipse}). The grey dashed line depicts the value observed for RX J1301.}
\label{fig:fs}
\end{figure}

\section{Binary system limits on RX J1301}
\label{sec:binary}

In this Section, we explore the limits we can put on the binary parameters if the binary BH hypothesis is to reproduce the observed QPEs. We first calculate the minimum eccentricity required to explain the ratio between long and short waiting times of consecutive flares (Section \ref{sec:ecc}). We then place limits on the binary parameters by comparing the observed long term evolution of the QPEs with the expected orbital evolution due to GW emission (Section \ref{sec:evolution}). RX J1301 is by far the most useful object for both of these tests.

\subsection{Eccentricity}
\label{sec:ecc}

For both GSN 069 and RX J1301, the waiting time from large flare to small flare is greater than the waiting time from small to large flare. This can be explained within the lensing model if the orbit is eccentric, since in this case the orbital angular velocity changes with orbital phase. According to the coordinate system defined in Fig. \ref{fig:ellipse}, the small flare (secondary is the lens) occurs when the orbital phase is equal to the viewer azimuth, $\phi_*=\phi_i$, and the large flare (primary is the lens) occurs when $\phi_*=\phi_i+\pi$. The orbital angular velocity takes its maximum value at peri-center ($\phi_*=0$). Therefore the lensing model requires that both QPE sources happen to have $\pi < \phi_i < 2\pi$, but that sources discovered in the future are equally likely to display the opposite behaviour of the longer waiting time being after the small flare (which would result from $0 < \phi_i < \pi$). 

In the 2019 observations of RX J1301, the short waiting time is $\sim 13$ ks, which is $\sim 40\%$ of the orbit. We can determine the minimum orbital eccentricity required to explain this by first integrating the Keplerian expression for orbital angular velocity to get the waiting time from peri-center to orbital phase $\phi_*$
\begin{equation}
    t(\phi_*) = \frac{P}{2\pi} \left\{ 2 \arctan\left[ \sqrt{\frac{1-\epsilon}{1+\epsilon}} \tan\left(\frac{\phi_*}{2}\right) \right] - \frac{\epsilon \sqrt{1-\epsilon^2}\sin\phi_*}{1+\epsilon \cos\phi_*} \right\}.
\end{equation}
The fraction of the orbit taken up by the short waiting time is then
\begin{equation}
    f_s(\epsilon,\phi_i) = \frac{ t(\phi_*=\phi_i+\pi) - t(\phi_*=\phi_i)  }{P}.
\end{equation}
Fig. \ref{fig:fs} shows $f_s$ plotted as a function of eccentricity for a range of $\phi_i$ values. The eccentricity for which this fraction takes the observed value $f_s=0.4$ (grey dashed line) depends on $\phi_i$. As can be seen in the plot, the minimum eccentricity for which $f_s=0.4$ occurs is for $\phi_i=-\pi/2$, since in this case the interval between the small and large flare is exactly the portion of the orbit with the fastest orbital angular velocity. Therefore the minimum eccentricity can be found by solving the equation $f_s(\epsilon_\mathrm{min},\phi_i=-\pi/2)=0.4$, yielding $\epsilon_\mathrm{min}\approx 0.16$. The eccentricity of RX J1301 in 2019, according to the lensing model, is therefore $0.16 \lesssim \epsilon \lesssim 1$. No formal limits can be placed on the eccentricity of GSN 069 because the large and small flares are almost equally spaced. This is most likely caused by the orbit being nearly circular, but is formally compatible with any eccentricity as long as $\phi_i \sim 0$ or $\phi_i \sim \pi$.

\subsection{Orbital evolution}
\label{sec:evolution}

We can also place limits on the binary parameters by comparing the observed long term evolution of the QPEs with the expected orbital evolution due to GW emission. Again, RX J1301 is the most useful object for this test, because there is a $\sim 19$ year baseline between the observation in 2000 and the recent observations.

The orbital separation and eccentricity will be reduced by energy lost through GW emission. The rate of change of the semi-latus rectum, $p$, and the eccentricity are respectively \citep{Peters1964,Gopakumar1997,Martel1999}
\begin{equation}
    \frac{dp}{dt} = -\frac{64}{5} \frac{(1-\epsilon^2)^{3/2}}{(p/r_g)^3} \left( 1 + \frac{7}{8}\epsilon^2 \right) \frac{M_1 M_2}{M^2}c,
\end{equation}
and
\begin{equation}
\frac{d\epsilon}{dt} = -\frac{304}{15} \frac{(1-\epsilon^2)^{3/2}}{(p/r_g)^4} \epsilon \left( 1 + \frac{121}{304}\epsilon^2 \right) \frac{M_1 M_2}{M^2} \frac{c}{r_g},
\end{equation}
where $r_g=GM/c^2$ is the gravitational radius of the binary system, $M=M_1+M_2$ and $c$ is the speed of light in a vacuum. We set an initial period and eccentricity, $P_0$ and $\epsilon_0$, and numerically evolve the above equations forwards or backwards in time from $t=0$. The period and semi-latus rectum are related by Kepler's law
\begin{equation}
    P=\frac{2\pi}{(GM)^{1/2}} \left( \frac{p}{1-\epsilon^2} \right)^{3/2}.
\end{equation}
We verify our scheme by comparing our numerical solution of the time-dependent period for $\epsilon=0$ with the analytic solution for period in the case of circular motion.

\begin{figure}
\centering
\includegraphics[width=8cm,trim=1.0cm 2.0cm 2.0cm 11.0cm,clip=true]{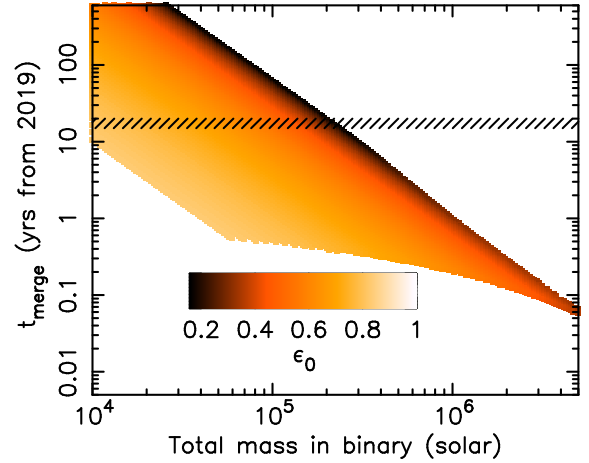}
\caption{Future evolution of RX J1301 from 2019 in the binary hypothesis. Each combination of binary mass $M$ and today's value of eccentricity $\epsilon_0$ that is consistent with the evolution of spacing between QPE flares from 2000 to 2019 is plotted as a coloured square (colour represents $\epsilon_0$, see inset legend). The hatched markings depict the nominal mission lifetime of \textit{LISA}.}
\label{fig:tmerge}
\end{figure}

\begin{figure}
\centering
\includegraphics[width=8cm,trim=1.0cm 1.5cm 2.0cm 2.5cm,clip=true]{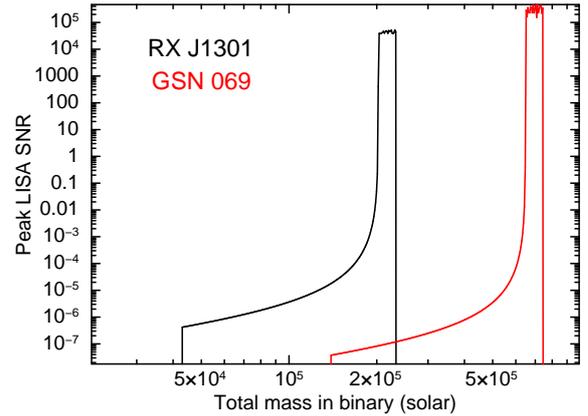}
\caption{Peak SNR for measurements of GW strain from RX J1301 (black) and GSN 069 (red) that can be made by \textit{LISA} during its operational lifetime (assumed to be 2034-2038) as a function of binary mass. We assume $\epsilon_0=0.16$ and $\epsilon_0=0$ for RX J1301 and GSN 069 respectively. The longer inferred orbital period of GSN 069 places it further away from merger for a given binary mass and therefore the binary must be more massive for it to merge during the \textit{LISA} mission.}
\label{fig:snrmax}
\end{figure}

For the case of RX J1301, the short and long waiting times between flares in 2019 are respectively $\sim 13$ ks and $\sim 20$ ks. The 2000 observation only contains one full flare, but the tail end of the previous flare can be seen at the start of the observation $\sim 18$ ks before the full flare. If the waiting time between flares in the 2000 observation was the short waiting time, then the short waiting time has reduced from $\sim 18$ ks to $\sim 13$ ks in $\sim 19$ yrs. We explore the allowed binary parameters by setting $P_0=33$ ks (the 2019 value) and trialling a range of present day eccentricities $\epsilon_0 \ge 0.16$ and a range of binary system mass values, $M$ (we set $q=1$). For each combination of mass and eccentricity trialled, we evolve the binary system backwards $19$ yrs to obtain the period and eccentricity in 2000. The resulting waiting time between flares in 2000 is then also dependent on the viewer azimuth, $\phi_i$. The 2019 value of $\phi_i$ is fixed by the observed $f_s$ for a particular trail $\epsilon_0$ value, but $\phi_i$ will have \textit{changed} since 2000 due to periastron precession. The precession period is $\sim (2/3) \pi (r_a/r_g) (1-\epsilon^2) P \sim 1$ yr \citep[e.g.][]{ShapiroKey2011}, and depends on a number of unknown quantities such as the spin parameters and spin alignment of the two BHs. The uncertainty in the periastron precession rate is therefore large enough for $\phi_i$ in 2000 to be completely unknown. Therefore, the values of $P$ and $\epsilon$ obtained by evolving the binary system backwards give rise to a \textit{range} of possible short waiting times, for $\phi_i$ in the range $0 < \phi_i < 2\pi$ (we do not see the peak of the first flare in 2000, therefore we do not know if it is larger or smaller than the full flare, therefore we do not know if $\phi_i <\pi$ or $\phi_i>\pi$). If the maximum possible short waiting time is less than $18$ ks, or if the minimum possible waiting time is greater than $18$ ks, then we discard the corresponding combination of $\epsilon_0$ and $M$.

We then evolve all of the successful combinations of $\epsilon_0$ and $M$ forward from 2019 for $600$ yrs and determine if and when the BHs will merge. Fig \ref{fig:tmerge} displays the results. All points with a coloured square have a solution that can explain the evolution of RX J1301 from 2000 to 2019, and the colour of the square represents the eccentricity. Since eccentricity reduces with time, and the binary is old enough to have a small separation, we see the lower eccentricity solutions (dark squares) as the more plausible. {We see that, as expected, lower mass binaries merge later. We only evolve the systems forward $600$ yrs, but extrapolating the trend from the figure gives $t_{\rm merge} \sim 2500$ yrs for $M=10^4~M_\odot$.} The hatched markings mark the nominal mission lifetime of the {\textit{Laser Interferometer Space Antenna} (\textit{LISA}; \citealt{Amaro-Seoane2017})} mission. We see that a fraction of our solutions merge during the lifetime of the \textit{LISA} mission. {\textit{LISA} may also be sensitive to GWs from the binary before merger. In Fig. \ref{fig:snrmax}, we explore this possibility by taking the (most likely) $\epsilon_0=0.16$ solutions and evolving them forwards in time, calculating the characteristic GW strain as
$h_c = 4 \pi^{2/3} (G M_{\rm ch})^{5/3} c^{-4} D^{-1} f^{2/3}$, where $f=2/P$ is the GW frequency, $D$ is the distance to the source (which we calculate from the redshift $z=0.023$ assuming $H_0=74$ km/s/Mpc) and $M_{\rm ch}=(M_1^3 M_2^3/M)^{1/5}$ is the chirp mass \citep[e.g.][]{Misner1973}. We then calculate the \textit{LISA} signal to noise ratio (SNR) by dividing by the \textit{LISA} noise \citep[][their equation 1]{Robson2019} at the relevant GW frequency. We plot the peak value that this SNR reaches during the nominal 2034-2038 \textit{LISA} lifetime as a function of binary mass (black line). We see that \textit{LISA} will detect the binary for only a limited range of binary mass, for which the merger occurs during or shortly after the \textit{LISA} mission. We repeat the same analysis for GSN 069 (red line), this time with $\epsilon_0=0$, $P_0=18$ hrs and $z=0.018$, and find that \textit{LISA} will be able to detect the binary for a similarly narrow range of binary mass.}

We do not yet have a long enough observational baseline to analyse the past evolution of the GSN 069 flares, but clearly future monitoring of the flare arrival times for both objects will enable us to better assess the feasibility of both being binary BH systems, and to determine the binary mass and therefore expected merger time. Useful constraints can be obtained with a baseline of a few years if flare arrival times relative to the flares in 2019 are measured as opposed to waiting times between consecutive flares. Our model predicts the relative spacings between small and large flares to change over a reasonably short timescale due to periastron precession (period of $\sim 1$ yr for RX J1301). Dense monitoring of the flare arrival times would therefore yield a measurement of the periastron precession rate. Finally, if the flares stop, it could be either because the two BHs have merged, or because the binary has became close enough to tidal forces to disrupt the mini-discs \citep{Haiman2017,Tang2018}.

\section{Modelling the GSN 069 QPEs as lensing flares}
\label{sec:profiles}

We now calculate self-lensing flares for a high inclination binary supermassive BH (SMBH) system in order to compare with the QPE flares observed from GSN 069. We first briefly investigate the properties of the GSN 069 flares, we then
describe the model,
before exploring the flare profiles produced by the model,
and we finally explore a large range of parameter space.

\subsection{The GSN 069 flares}
\label{sec:data}

\begin{figure}
\centering
\includegraphics[width=8cm,trim=1.0cm 1.5cm 2.0cm 2.5cm,clip=true]{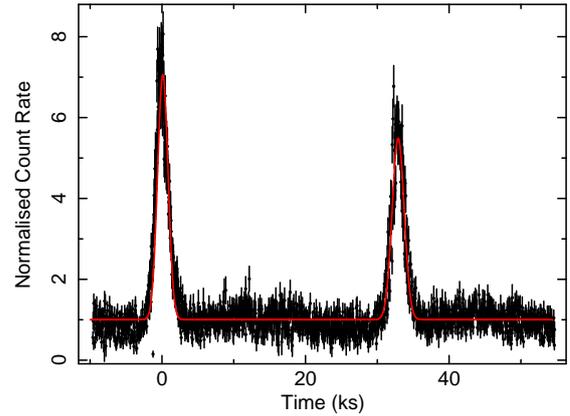}
\caption{0.4-2 keV \textit{XMM-Newton} light curve of GSN 069 from $16^{\rm th}$ January 2019 folded on a period of $64.3555$ ks and normalised to the quiescent level (black). The light curve includes three large flares and two small flares. We model the folded light curve with a constant (fixed to unity) plus two Gaussian components (red). The first Gaussian component represents the large flare (centroid=$0.1258$ ks, width $\sigma=0.722$ ks, normalisation $N=6.055$) and the second represents the small flare (centroid=$32.907$ ks, $\sigma=0.835$ ks, $N=4.490$). The waiting time from large flare to small flare is therefore $\sim 49\%$ of the period.}
\label{fig:gsn069}
\end{figure}

Here, our primary aim is to roughly determine basic properties of the GSN 069 flares that we aim to reproduce with our self-lensing model. We leave a comprehensive timing analysis to a later paper (Miniutti et al in prep). Fig. \ref{fig:gsn069} shows a $0.4-2$ keV light curve from GSN 069 folded on a period of $64.3555$ ks and normalised to the quiescent level (black points). The large and small flares are clearly visible. We fit the flares with Gaussian functions and the quiescent flux with a constant frozen to unity (red lines). The best fitting Gaussian parameters (quoted in the caption) indicate that the large and small flares have respectively a peak magnification of 7.055 and 5.49 and a full width at half maximum (FWHM) of 1.35 ks and 1.62 ks\footnote{Note that our model for each flare is a Gaussian function plus a unity quiescent level, so the peak magnification is $1+N$ and the FWHM is $2\sigma \sqrt{2 \ln[2N/(N-1)]}$}. The waiting time from small flare to large flare is $\sim 49\%$ of the period, compared with $\sim 40\%$ for RX J1301. Our analysis here only considers the longest public \textit{XMM-Newton} observation of the source (from January 2019), which includes three large flares and two small flares. We see that the period appears to be stable over this observation. Overall, there are 15 flares in publicly available data, distributed over three \textit{XMM-Newton} observations and one Chandra observation, covering a duration of $\sim 5$ months. The period is less stable over this longer timescale, displaying small but noisy timing residuals compared to a model with a constant or smoothly evolving period (Miniutti et al in prep). The flare amplitude increases dramatically with energy band (see Figs 2 and 3 of \citealt{Miniutti2019}).

\subsection{Self-lensing model}
\label{sec:model}

\begin{figure}
\centering
\includegraphics[width=8cm,trim=5.5cm 8.5cm 4.8cm 4.0cm,clip=true]{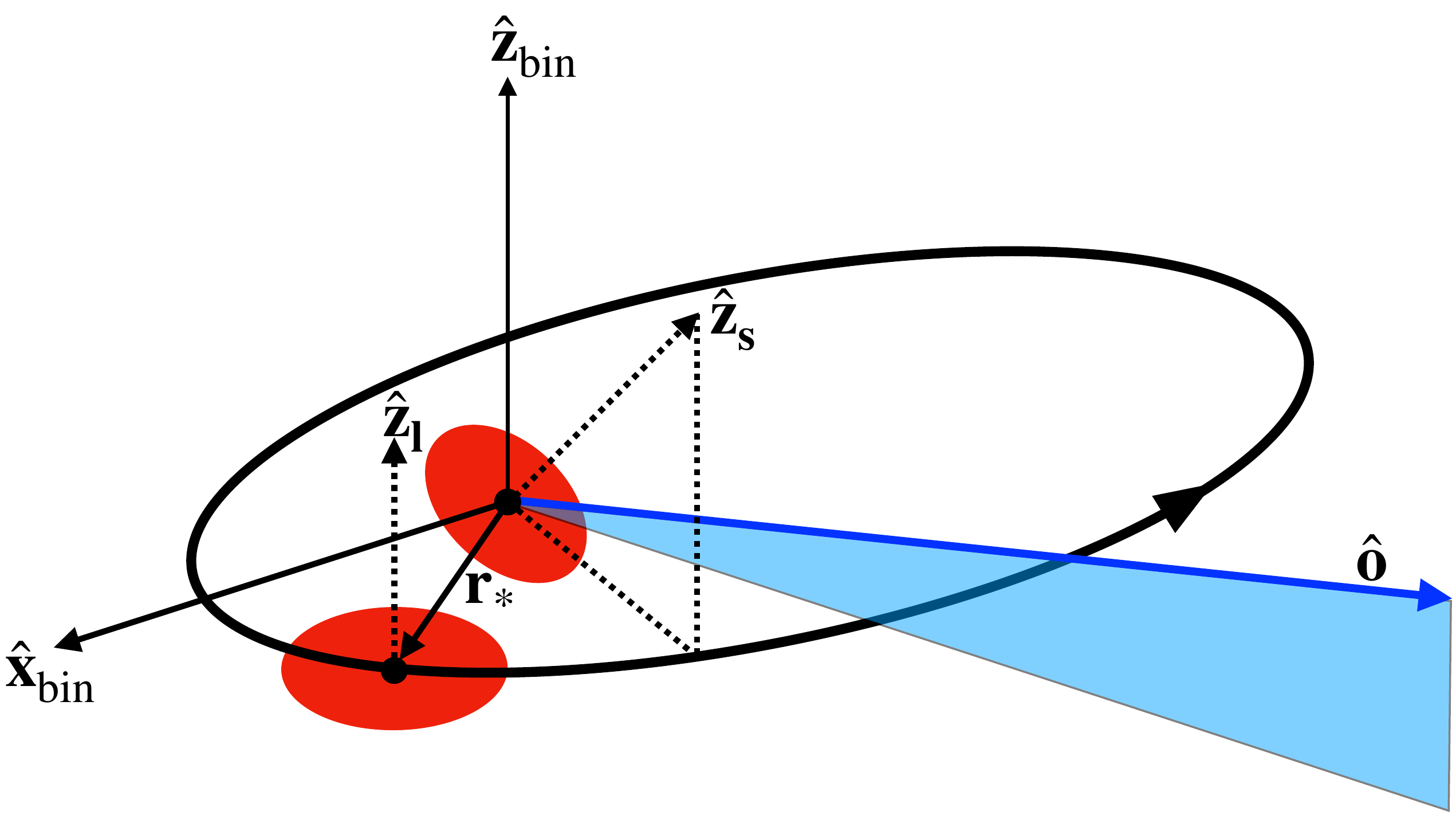}
\caption{Schematic of our model geometry. A mini-disc (red) surrounds each BH (black circles). The primary is at the origin and the secondary follows an anti-clockwise elliptical orbit in the plane perpendicular to $\mathbf{\hat{z}_{\rm bin}}$, with a focus at the origin and pericentre when the vector connecting the two BHs, $\mathbf{r_*}$, aligns with $\mathbf{\hat{x}_{\rm bin}}$. The observer's line of sight is represented by the vector $\mathbf{\hat{o}}$ (blue). The rotation axis of the background and foreground mini-discs are respectively $\mathbf{\hat{z}_s}$ and $\mathbf{\hat{z}_l}$. In the case pictured, the background disc is tilted somewhat towards and to the left of the observer's line of sight and the foreground disc aligns with the binary rotation axis.}
\label{fig:schem}
\end{figure}

Fig. \ref{fig:schem} illustrates the geometry of our model. The source (background) BH, mass $M_s$, is pictured at the origin, with the lensing (foreground) BH, mass $M_\ell$, orbiting it. We define the mass ratio as $q=M_l/M_s$. The orbit is in general elliptical with semi-major axis $r_a$, although we go on to only explore circular orbits in the following sub-section. The binary orbital axis is marked as $\mathbf{\hat{z}_{\rm bin}}$, and peri-center is marked by the vector $\mathbf{\hat{x}_{\rm bin}}$. The vector $\mathbf{\hat{o}}$ points from the source BH to the observer, such that the inclination angle $i$ is the angle between $\mathbf{\hat{z}_{\rm bin}}$ and $\mathbf{\hat{o}}$. The viewer azimuth, $\phi_i$, is the angle between $\mathbf{\hat{x}_{\rm bin}}$ and the projection of $\mathbf{\hat{o}}$ onto the binary plane. For circular orbits, we simply choose $\phi_i=0$. The vector $\mathbf{r_*}$ points from the source BH to the lens BH, such that its orientation moves anti-clockwise around the binary plane as the orbital phase $\phi_*$ increases.

Each BH has an associated mini-disc. The vectors $\mathbf{\hat{z}_s}$ and $\mathbf{\hat{z}_l}$ denote the rotational axis for the background and foreground mini-discs respectively. These vectors are defined by the angles $\delta_s$, $\gamma_s$, $\delta_l$ and $\gamma_l$, where $\cos\delta_s=\mathbf{\hat{z}_{\rm bin}} \cdot \mathbf{\hat{z}_s}$, $\gamma_s$ is the angle between the projections of $\mathbf{\hat{z}_s}$ and $\mathbf{\hat{o}}$ on the binary plane, and $\delta_l$ and $\gamma_l$ are defined similarly. The $\delta$ angles are therefore polar angles, and the $\gamma$ angles are azimuthal angles (in a right-handed coordinate system). The mathematical details of our coordinate system are given in Appendix (\ref{sec:coord}).

We make the following assumptions about the geometry of the system:
\begin{itemize}
    \item {The mini-discs are both stable over many binary orbits. This may not be the case for a particularly compact ($r_a \lesssim 60~r_g$) binary \citep{Tang2018,Paschalidis2021}.}
    \item Each mini-disc is aligned with the corresponding BH spin axis. This may not be true in general, and a misalignment does influence the dynamics \citep{Bardeen1975,Liska2019b} and the observed spectrum \citep{Abarr2020} of the disc, but we make this simplifying assumption in order to keep the problem tractable.
    \item The two BHs have the same dimensionless spin parameter, $a$, as one another. Again this is a simplifying assumption to keep the number of model parameters down.
    \item The inner radius of each mini-disc is at the innermost stable circular orbit (ISCO), which decreases with increasing spin parameter $a$ \citep{Bardeen1972}. This is thought to be the case for BH X-ray binaries in the soft state \citep[e.g.][]{Done2007}, for which the thermal spectrum is very similar to that of the QPE sources.
    \item The outer edge of each mini-disc is $r_{\rm out}=0.25~r_a$. This is set by the dynamics of the binary system, including tidal forces and resonances (e.g. \citealt{Artymowicz1994,Farris2014}). We adopt the value approximately found by \citet{Bowen2017} in their relativistic simulations of a $q \sim 1$ binary system (see their Table 3).
    \item The mini-discs are both circular and have well-defined outer edges. \citet{Bowen2017} note that this is not strictly the case in their simulations, with the outer edge on the near side of the other BH differing from that on the far side. We also ignore the streams of material flowing from the circumbinary disc to the mini-discs, and between the two mini-discs.
    \item The discs each radiate a multi-temperature blackbody spectrum with a temperature profile $T(r) \propto r^{-3/4} [1-\sqrt{r/r_{\rm in}}]$, which corresponds to gravitational energy release with zero torque at the inner boundary \citep{Shakura1973,Novikov1973}. The peak disc temperature is a model parameter. We fix it to $kT_{\rm max} = 50$ eV to match the spectrum of GSN 069 \citep{Miniutti2019}.
\end{itemize}

\begin{figure}
\centering
\includegraphics[width=\columnwidth,trim=1.5cm 2.0cm 1.5cm 10.5cm,clip=true]{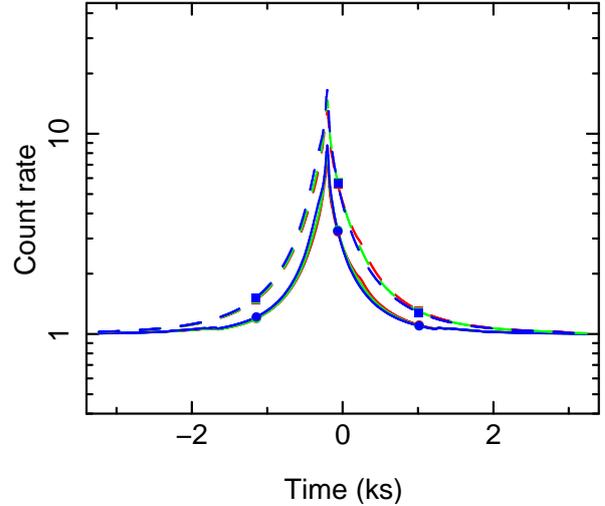}
\caption{Light curves calculated for model parameters: $a=0$, $\delta=0^\circ$, $i=89.5^\circ$, $M=10^6~M_\odot$. Red, green and blue lines represent the $0.2-0.3$ keV, $0.4-0.5$ keV and $0.8-1.0$ keV \textit{XMM-Newton} count rate normalised to the quiescent level. Dashed lines ignore occultation by the foreground mini-disc and solid lines account for it. Filled circles and squares denote the times for which images are plotted in Fig. \ref{fig:image_a0_d0}.}
\label{fig:lc_a0_d0}
\end{figure}

\begin{figure}
\centering
\includegraphics[width=\columnwidth,trim=1.0cm 1.5cm 3.0cm 3.0cm,clip=true]{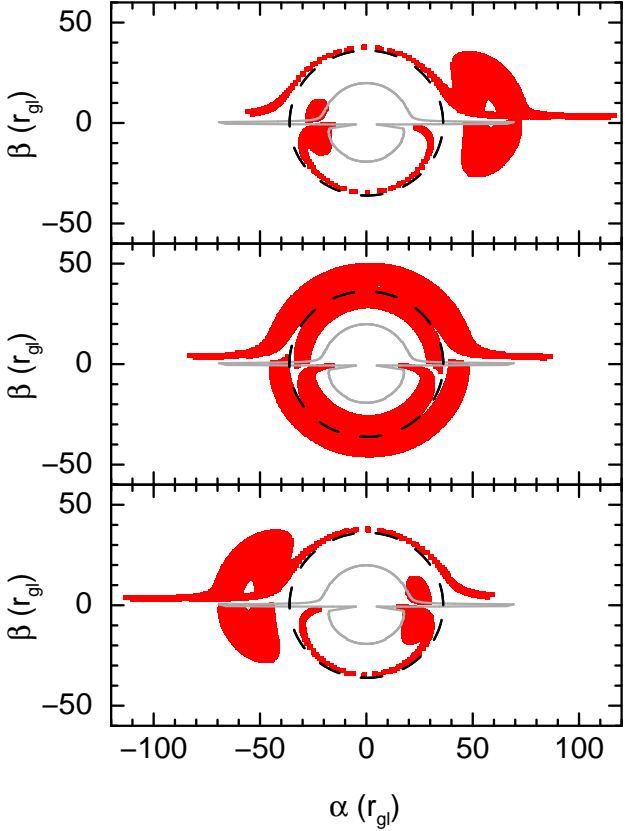}
\caption{Images at three points in time denoted by the filled circles in Fig. \ref{fig:lc_a0_d0}. Model parameters are: $a=0$, $\delta=0^\circ$, $i=89.5^\circ$, $M=10^6~M_\odot$. The background mini-disc is shown in red, the outline of the foreground mini-disc is in grey and the black dashed line denotes the Einstein radius of the foreground BH, which is always at the origin. The image is orientated such that the rotation axis of the foreground mini-disc is always vertical, and distances are expressed in units of $r_{g\ell}=GM_\ell/c^2$. The background mini-disc is warped by lensing around both the background and foreground BHs.}
\label{fig:image_a0_d0}
\end{figure}

\begin{figure}
\centering
\includegraphics[width=\columnwidth,trim=1.5cm 2.0cm 1.5cm 10.5cm,clip=true]{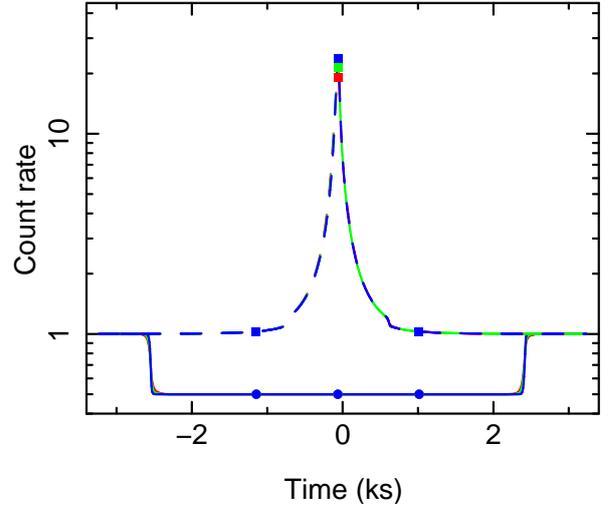}
\caption{Light curves calculated for model parameters: $a=0$, $\delta=30^\circ$, $\gamma=0^\circ$, $i=89.9^\circ$, $M=10^5~M_\odot$. We use the same plotting conventions as Fig. \ref{fig:lc_a0_d0}. For these parameters, occultation by the foreground mini-disc leads to an eclipse (solid lines). Filled circles denote the times for which images are plotted in Fig. \ref{fig:image_a0_d30}.}
\label{fig:lc_a0_d30}
\end{figure}

\begin{figure}
\centering
\includegraphics[width=\columnwidth,trim=1.0cm 1.5cm 3.0cm 3.0cm,clip=true]{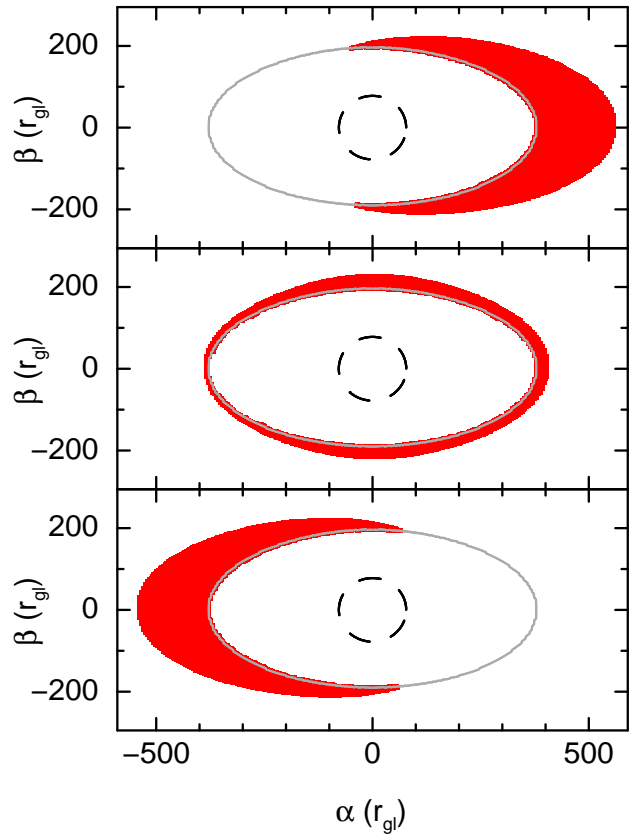}
\caption{Images at three points in time denoted by the filled circles in Fig. \ref{fig:lc_a0_d30}. Model parameters are: $a=0$, $\delta=30^\circ$, $\gamma=0^\circ$, $i=89.9^\circ$, $M=10^5~M_\odot$. We use the same plotting conventions as in Fig. \ref{fig:image_a0_d0}.}
\label{fig:image_a0_d30}
\end{figure}

We first ray trace each disc individually in the Kerr metric. For each mini-disc, we define an image plane perpendicular to the observer's line of sight, $\mathbf{\hat{o}}$, with the vertical axis $\mathbf{\hat{\beta}}$ parallel to the projection of the disc spin axis on the image plane. The horizontal axis $\mathbf{\hat{\alpha}}$ is perpendicular to both $\mathbf{\hat{o}}$ and $\mathbf{\hat{\beta}}$, forming a right-handed coordinate system. The horizontal and vertical parameters on the image plane, $\alpha$ and $\beta$ are the impact parameters at infinity, and the BH is at the origin ($\alpha=\beta=0$). Each point on the image plane, represented by the vector $\mathbf{b} = \alpha~\mathbf{\hat{\alpha}} + \beta~\mathbf{\hat{\beta}}$, corresponds to a unique null-geodesic in the Kerr metric. We trace rays backwards from a grid of $(\alpha,\beta)$ coordinates towards each BH, solving for the radius $r$ that the geodesic crosses the equatorial plane of the BH. We use the publicly available code \textsc{ynogk}, \citep{Wang2012} which is based on the code \textsc{geokerr} \citep{Dexter2009}\footnote{The ray tracing code we used is also publicly available as part of the \textsc{reltrans} X-ray reverberation mapping code \citep{Ingram2019}.}.

We then account for lensing of the background mini-disc by the foreground BH. This leads to each patch of the background mini-disc having two images: one inside (the internal image) and one outside (the external image) of the Einstein radius of the foreground BH, $r_E$. The position of the internal and external images on the observer plane are given respectively by the vectors $\mathbf{b_-}$ and $\mathbf{b_+}$. For simplicity, we ignore strong field effects, which is not necessarily a good assumption, but provides a good starting point to determine whether or not the model qualitatively works. In this case
\begin{equation}
    \mathbf{b_{\pm}} = \mathbf{b_*} + \frac{1}{2} \left[ 1 \pm \sqrt{1 + \left( \frac{2}{u} \right)^2 } \right] ( \mathbf{b} - \mathbf{b_*} ),
\end{equation}
where $u = |\mathbf{b} - \mathbf{b_*}| / r_E$ and $\mathbf{b_*}$ is the projection of $\mathbf{r_*}$ on the image plane \citep{Paczynski1996,Agol2002,D'Orazio2018}. The apparent area of each image is multiplied by a magnification factor
\begin{equation}
    \mathcal{M}_\pm = \frac{1}{2} \left[ \frac{u^2+2}{u\sqrt{u^2+4}} \pm 1 \right].
    \label{eqn:mag}
\end{equation}
The Einstein radius is given by $r_E = \sqrt{4 r_{g\ell}  \mathbf{r_*} \cdot \mathbf{\hat{o}}}$, where $r_{g\ell}=GM_\ell/c^2$. We account for Shapiro delays -- caused by geodesics that are more heavily distorted by gravitational lensing having an increased path length -- with a simplified prescription described in Appendix (\ref{sec:shap}).

We calculate the time dependent observed spectrum of the background mini-disc for a grid of time bins, and the spectrum of the foreground mini-disc (details in Appendix \ref{sec:spec}). We account for interstellar absorption using the \textsc{xspec} model \textsc{tbabs} \citep{Wilms2000}. We fix the hydrogen column density to $N_h=5\times 10^{20} \mathrm{cm}^{-2}$, to match GSN 069 \citep{Miniutti2019}. We then fold the spectrum of the background mini-disc for each time bin, and also the constant spectrum of the foreground mini-disc, around the \textit{XMM-Newton} EPIC-pn response matrix, and integrate the resulting counts spectra over the specified range of energy channels to create a light curve. We then add on the constant count rate of the foreground mini-disc and finally normalise to the quiescent level. {Note that it is necessary to account for interstellar absorption because the response of \textit{XMM-Newton} is not diagonal, especially in soft X-rays. Absorption can therefore lead to the soft X-ray energy channels being dominated by `mis-classified' higher energy photons. If the flare profile depends on energy, absorption will therefore influence the count rate as a function of time in a given energy channel, even on timescales on which $N_H$ is constant \citep{Ingram2019}.}

{We do not include Doppler modulation due to centre-of-mass motion of each mini-disc around the binary centre of mass. This does not contribute to the flare profile since the orbital motion of each BH is entirely transverse during the flares. The orbital motion should, however, contribute a periodic Doppler modulation outside of the flares \citep[e.g.][]{D'Orazio2018,Hu2020}. For a $\sim$unity mass ratio, this modulation will be fairly subtle, since one BH is approaching the observer when the other is receding from the observer and so the summed flux from the two mini-discs will not be modulated as strongly as for $M_1 \gg M_2$. However, even for $q=1$ there should still be a subtle Doppler modulation that peaks between flares since the approaching mini-disc should be boosted more strongly than the receding mini-disc is suppressed (specific flux is $\propto$ Doppler factor to the third power).}

\subsection{Predicted flare profiles}

We now explore the predicted flare profiles of our model. We consider the parameters relevant to GSN 069, specifically an orbital period of 18 hrs and a possible binary mass range of $\sim 5\times 10^4-5\times 10^{6}~M_\odot$
\footnote{Note that the \citet{Miniutti2019} spectroscopic mass estimate of $M\sim 4\times 10^5~M_\odot$ assumed a single disc inclined by $45^\circ$ that extends to the ISCO of a Schwarzschild BH. Assuming the presence of a second disc, a higher inclination angle or a higher BH spin would \textit{increase} the inferred mass, and so we see the range $M \gtrsim 4 \times 10^{5}~M_\odot$ as the most realistic, but the uncertainty is large enough to make it worth exploring significantly lower binary masses.}.
In order to reproduce the small difference between small and large QPE flares, we consider only binary parameters that would result in the two lensing flares per orbital period having the same profile. We therefore fix the mass ratio to $q=M_\ell/M_s=1$, the eccentricity to $\epsilon=0$, and set the misalignment angles of the two BHs equal to one another: $\delta_\ell=\delta_s=\delta$, $\gamma_\ell=\gamma_s=\gamma$. Although the large-flare-small-flare behaviour requires that $q$ cannot be exactly unity, we note that the large flares are only $\sim 30\%$ larger than the small flares (see Fig. \ref{fig:gsn069}) and so $q$ must be \textit{close} to unity. We only consider $q=1$ in this section, since we are looking to explain the general properties of the flares. Any departures from $q=1$ that make a significant difference to the flare profiles will also predict the large flare to be significantly larger than the small flare, which is not observed. We also note that the observed small noisy departures from a smoothly evolving period in the light curve can be explained as stochastic variations of the mini-disc structure, caused both by tidal forces and potentially relativistic precession.

Fig. \ref{fig:lc_a0_d0} shows the predicted light curve for an aligned system ($\delta=0$) of two Schwarzschild BHs with a total binary mass of $M=10^6 M_\odot$, viewed from an inclination angle of $i=89.5^\circ$. Red, green and blue lines represent respectively the $0.2-0.3$ keV, $0.4-0.5$ keV and $0.8-1.0$ keV light curves, normalised to the quiescent-level count rate. Occultation by the foreground mini-disc is taken into account for the solid lines and ignored for the dashed lines. We see that the flare amplitude is comparable to that observed for QPEs (Fig. \ref{fig:gsn069}), but the duration is shorter. The QPE flares from GSN 069 (and indeed RX J1301) have a very strong energy dependence, with larger flares in higher energy bands \citep{Miniutti2019}. Our model does predict stronger flares in higher energy bands because the Doppler blue/red shifted (approaching/receding) side of the disc is magnified during the rising/falling phase of the flare. However, the resulting energy dependence shown in Fig. \ref{fig:lc_a0_d0} is far more subtle than what is observed for QPEs.

Fig. \ref{fig:image_a0_d0} shows images of the system at three snapshots of time denoted by the solid circles in Fig. \ref{fig:lc_a0_d0}. The lensing (foreground) BH is at the origin, and it's Einstein radius is represented by the dashed black line. The background mini-disc is coloured red and the outline of the foreground mini-disc, which eclipses any part of the background mini-disc that it passes in front of, is shown in grey. Note that the $\sim r^{-3}$ radial emissivity of the background mini-disc is not shown in these images, but \textit{is} accounted for in our model. Distances are represented in units of $r_{g\ell}=GM_\ell/c^2$, and all images are oriented such that the rotation axis of the foreground mini-disc is vertical. We see that, from the top to the bottom panel, the background mini-disc moves from right to left, and is more heavily warped the closer on the image plane it is to the foreground BH. Each patch of the background mini-disc has an internal and an external image, one inside and one outside of the Einstein radius of the foreground BH. As a patch of the background mini-disc passes closer to the foreground BH, both images move closer to the Einstein radius and the magnification increases.

Occultation by the foreground mini-disc has little effect for the aligned case explored in Figs.
\ref{fig:lc_a0_d0} and 
\ref{fig:image_a0_d0} because the projected image of the foreground mini-disc is largely \textit{inside} of the Einstein radius. The foreground disc therefore cuts out almost none of the external image of the background disc, and only the lowest magnification part of the internal image. This is not the case for all parameter combinations.
Fig. \ref{fig:lc_a0_d30} shows the light curve for $a=0$, $i=89.9^\circ$, $\delta=30^\circ$, $\gamma=0^\circ$ and $M=10^5~M_\odot$. The solid line now displays an eclipse down to half of the quiescent level (since half of the total flux is from the foreground disc) with no lensing flare at all. As Fig.
\ref{fig:image_a0_d30}
demonstrates, this is because the projection of the foreground mini-disc is now larger than the Einstein radius of the foreground BH, and so all of the magnified geodesics are cut out. The larger projected size of the foreground disc, relative to the Einstein radius of the foreground BH, is due partly to its lower inclination to the observer's line of sight (i.e. $\beta=30^\circ$, $\gamma=0$ means that the mini-discs are viewed less edge-on than before). The lower binary mass is also important though because the disc outer radius, $r_{\rm out}$, is proportional to the binary separation, $r_a$, whereas the Einstein radius is $r_E \propto r_{g\ell}^{1/2} r_a^{1/2}$. For a given orbital period and mass ratio, Kepler's law therefore gives $r_E/r_{\rm out} \propto M^{1/3}$, and so occultation will matter less for higher mass systems.

\begin{figure}
\centering
\includegraphics[width=\columnwidth,trim=1.5cm 2.0cm 1.5cm 10.5cm,clip=true]{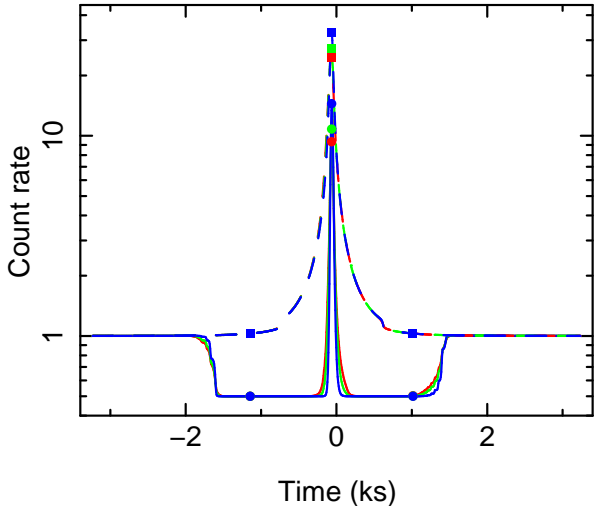}
\caption{Light curves calculated for model parameters: $a=0$, $\delta=10^\circ$, $\gamma=50^\circ$, $i=89.9^\circ$, $M=10^5~M_\odot$. We use the same plotting conventions as Fig. \ref{fig:lc_a0_d0}. For these parameters, the light curve features an eclipse and a sharp lensing flare (solid lines). Filled circles denote the times for which images are plotted in Fig. \ref{fig:image_a0_d10}.}
\label{fig:lc_a0_d10}
\end{figure}

\begin{figure}
\centering
\includegraphics[width=\columnwidth,trim=1.0cm 1.5cm 3.0cm 3.0cm,clip=true]{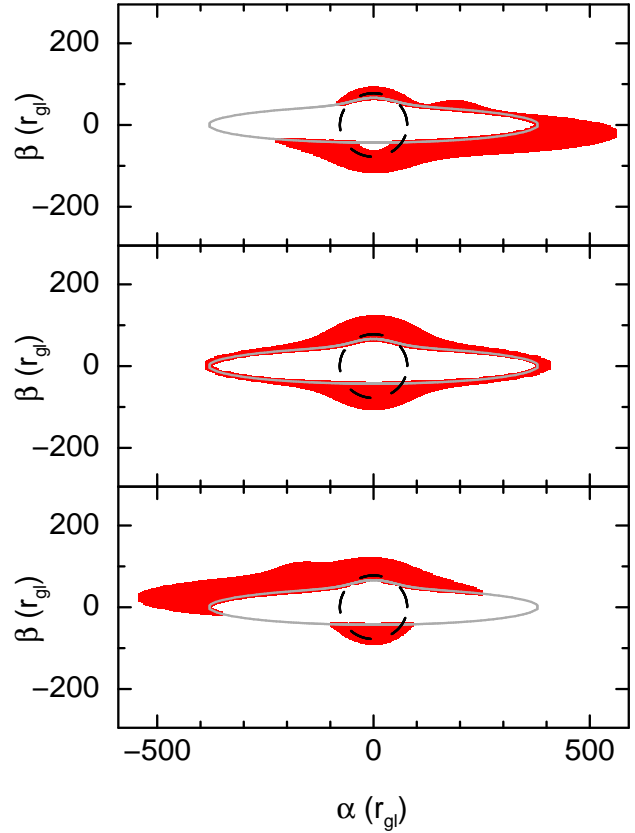}
\caption{
Images at three points in time denoted by the filled circles in Fig. \ref{fig:lc_a0_d10}. Model parameters are: $a=0$, $\delta=10^\circ$, $\gamma=50^\circ$, $i=89.9^\circ$, $M=10^5~M_\odot$. We use the same plotting conventions as in Fig. \ref{fig:image_a0_d0}. We see that the background mini-disc is lensed both around the background BH (the hump at $\alpha\sim 200~r_g$ in the top panel and $\sim -200~r_g$ in the bottom panel) and the foreground BH.}
\label{fig:image_a0_d10}
\end{figure}

There are also parameter combinations for which an eclipse with a sharp central lensing flare is predicted. Figs. \ref{fig:lc_a0_d10} and \ref{fig:image_a0_d10} show the light curve and images for such an example ($a=0$, $i=89.9^\circ$, $\delta=10^\circ$, $\gamma=50^\circ$ and $M=10^5~M_\odot$). In this case, we see from Fig. \ref{fig:image_a0_d10} that the foreground disc cuts out all but a small fraction of the internal image, but the part of the image that is not cut out is closest to the Einstein radius and is therefore highly magnified. The flare is therefore very sharp indeed, with a large amplitude.

\subsection{Parameter space exploration}

\begin{figure*}
\centering
\includegraphics[width=\columnwidth,trim=1.5cm 2.5cm 1.5cm 10.5cm,clip=true]{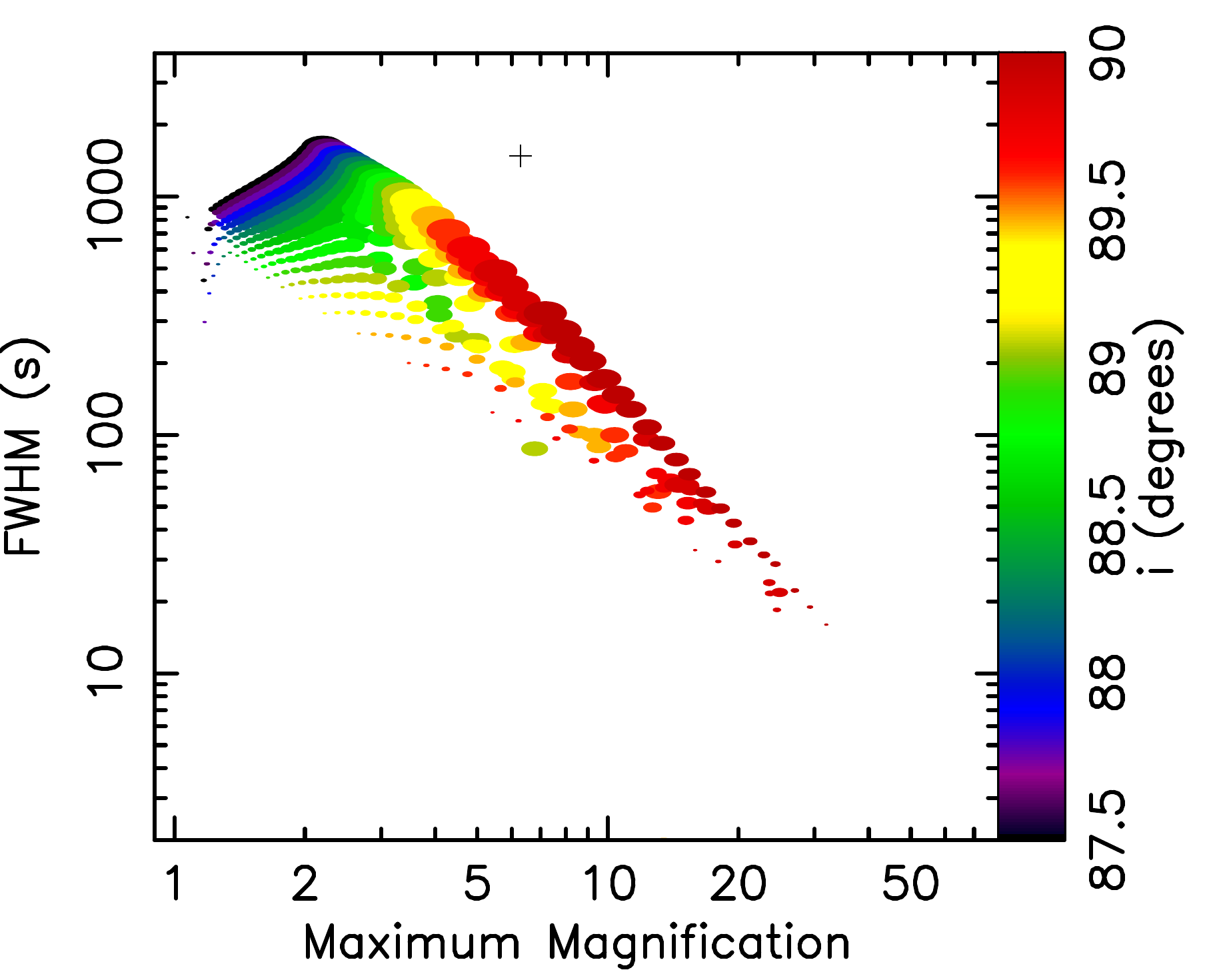}
\includegraphics[width=\columnwidth,trim=1.5cm 2.5cm 1.5cm 10.5cm,clip=true]{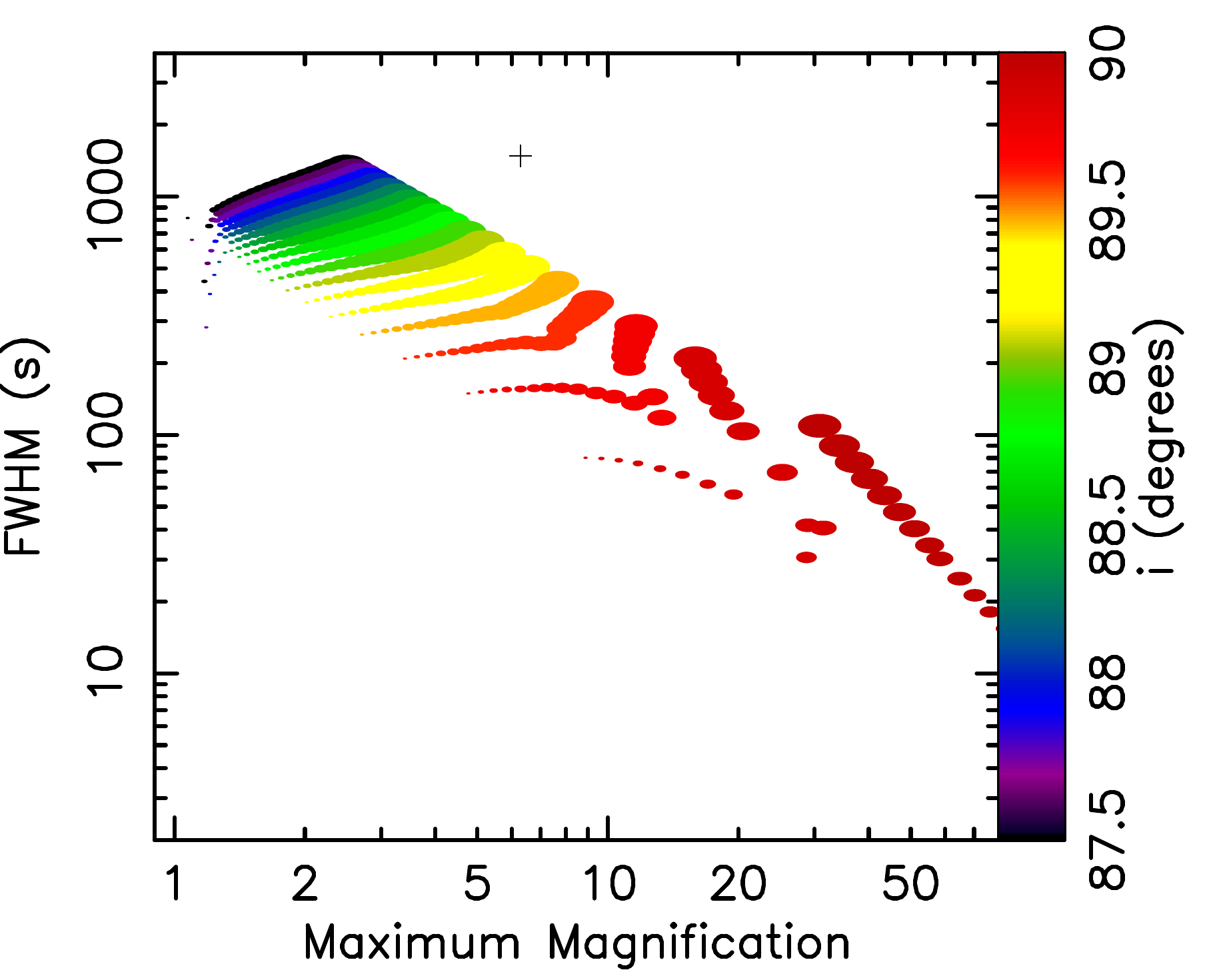}
\caption{Lensing flare FWHM versus peak magnification for an aligned ($\delta=\gamma=0^\circ$) binary system with spin parameter $a=0$ (left) and $a=0.998$ (right). Binary mass is varied in the range $M=5\times 10^4 - 5\times 10^6~M_\odot$, with larger markers corresponding to larger mass. Inclination angle is denoted by the colour code. The black cross corresponds to the QPEs in GSN 069.}
\label{fig:d0}
\end{figure*}

We now explore parameter space to determine whether our model can reproduce the light curve of GSN 069. We aim to investigate whether the model can reproduce the basic flare properties averaged over large and small flares when we set $q=1$; in the knowledge that any departure from unity mass ratio will naturally reproduce the observed large-flare-small-flare behaviour. From our fit to the folded light curve in Fig. \ref{fig:gsn069}, the average peak magnification is $\approx  6.3$ and the average FWHM is $\approx 1.5$ ks. We are interested both in whether the model can reproduce these properties at all and also how commonly it reproduces these values, which tells us the level of fine-tuning required for the model to explain the observed QPEs. Our model has five free parameters: $a$, $M$, $i$, $\delta$ and $\gamma$. This  represents a large parameter space to explore with a single grid, and so we explore a grid of $i$ and $M$ values for a few fixed combinations of $a$, $\delta$ and $\gamma$.

Fig. \ref{fig:d0} is for an aligned binary system ($\delta=\gamma=0^\circ$) with BH spin parameter $a=0$ (left) and $a=0.998$ (right). We trial a mass range of $M=5\times 10^4 - 5\times 10^6~M_\odot$, with larger markers representing larger mass values, and an inclination range of $i=87.5^\circ-90^\circ$ (colour coded). The black cross represents the QPEs observed from GSN 069. We see that the aligned model cannot produce lensing flares simultaneously as long and as bright as the GSN 069 QPEs. For a given binary mass, the flare peak and FWHM are anti-correlated with one another, and increasing the inclination increases the peak whilst decreasing the FWHM. Increasing the mass moves the trend line closer to the top-right of the diagram (i.e. closer to the peak and FWHM values of GSN 069) for the lower inclination angles, where as the mass dependence becomes more complex as the inclination angle approaches $90^\circ$. The flares tend to be more powerful for the case of high spin, since the ISCO is smaller and therefore the background source is more compact and so can be more efficiently lensed.

\begin{figure*}
\centering
\includegraphics[width=\columnwidth,trim=1.5cm 2.5cm 1.5cm 10.5cm,clip=true]{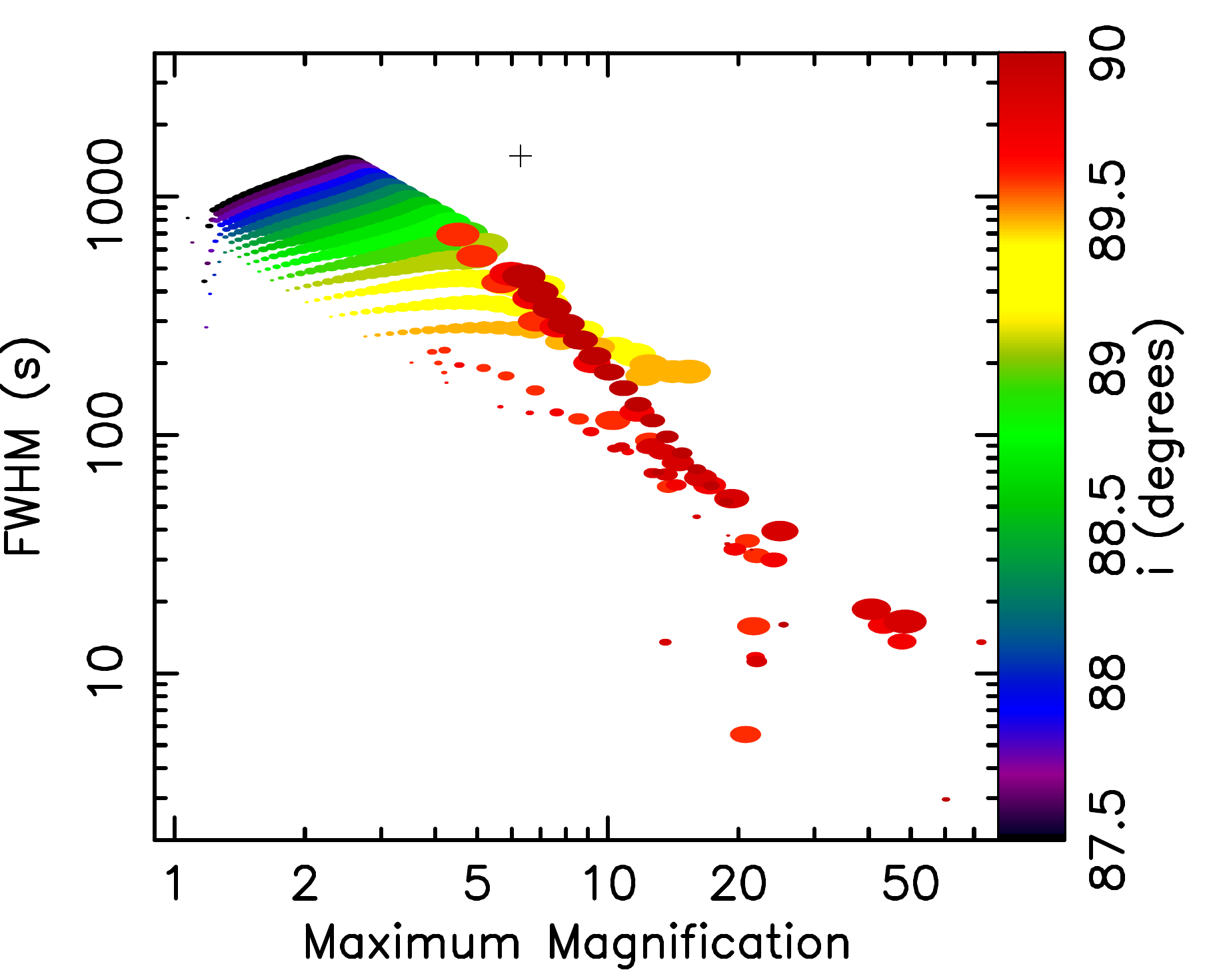}
\includegraphics[width=\columnwidth,trim=1.5cm 2.5cm 1.5cm 10.5cm,clip=true]{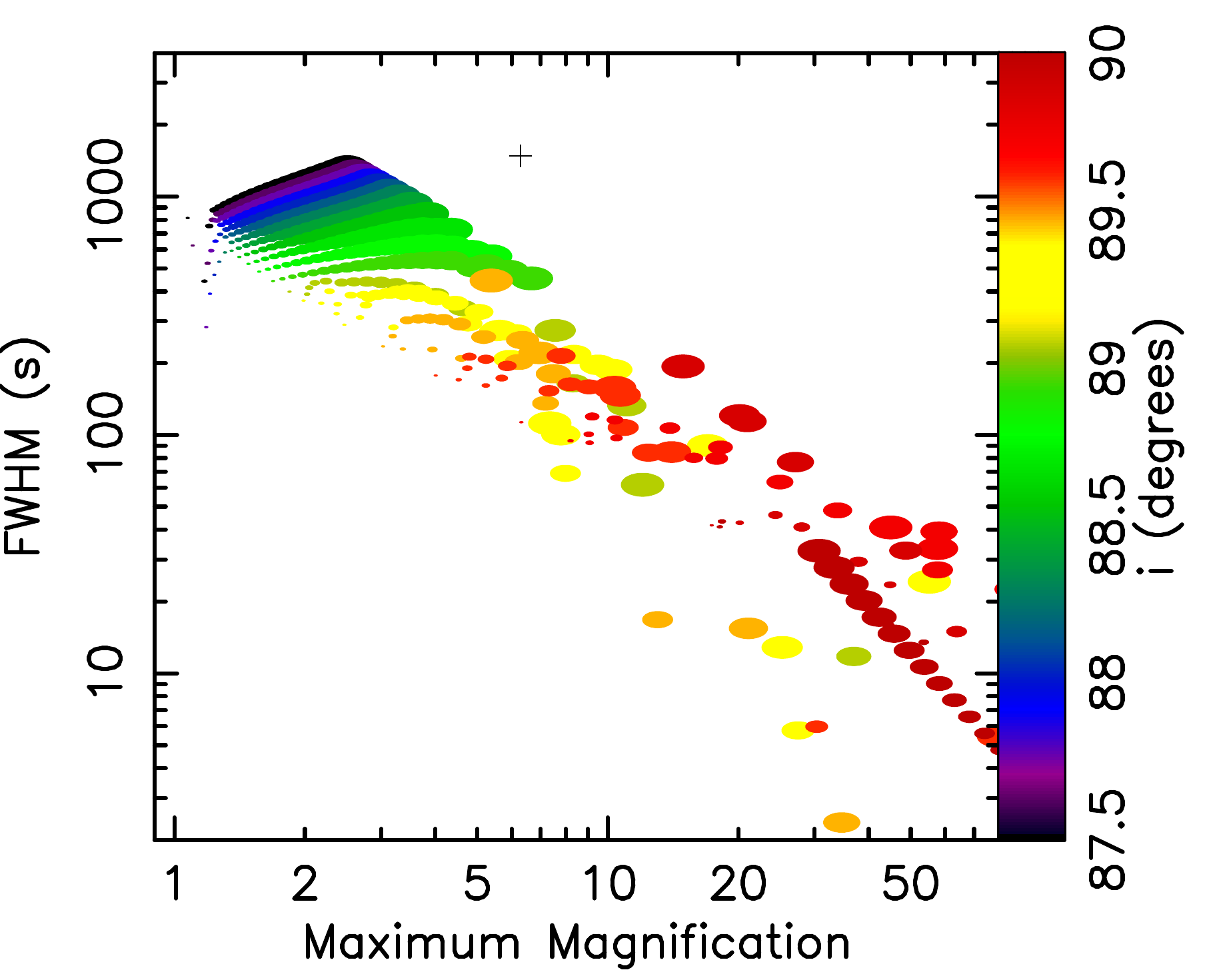}
\caption{Lensing flare FWHM versus peak magnification for two configurations in which we view the underside of the source and lens disc ($\gamma=180^\circ$). The BH spin parameter is $a=0.998$. Left: $\delta=0.5^\circ$; Right: $\delta=1^\circ$. Binary mass is varied in the range $M=5\times 10^4 - 5\times 10^6~M_\odot$, with larger markers corresponding to larger mass. Inclination angle is denoted by the colour code. The black cross corresponds to the QPEs in GSN 069.}
\label{fig:g180}
\end{figure*}

\begin{figure*}
\centering
\includegraphics[width=\columnwidth,trim=1.5cm 2.5cm 1.5cm 10.5cm,clip=true]{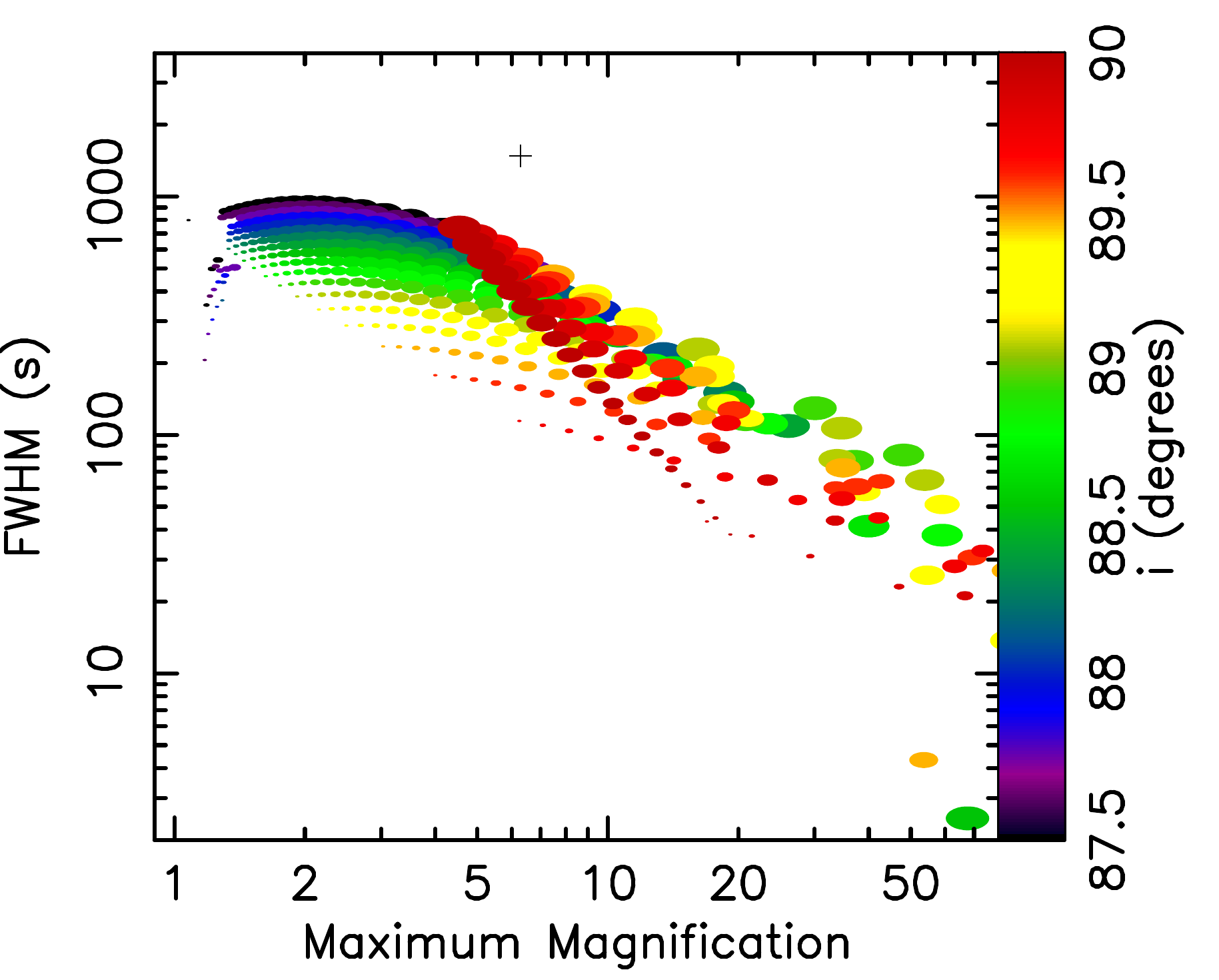}
\includegraphics[width=\columnwidth,trim=1.5cm 2.5cm 1.5cm 10.5cm,clip=true]{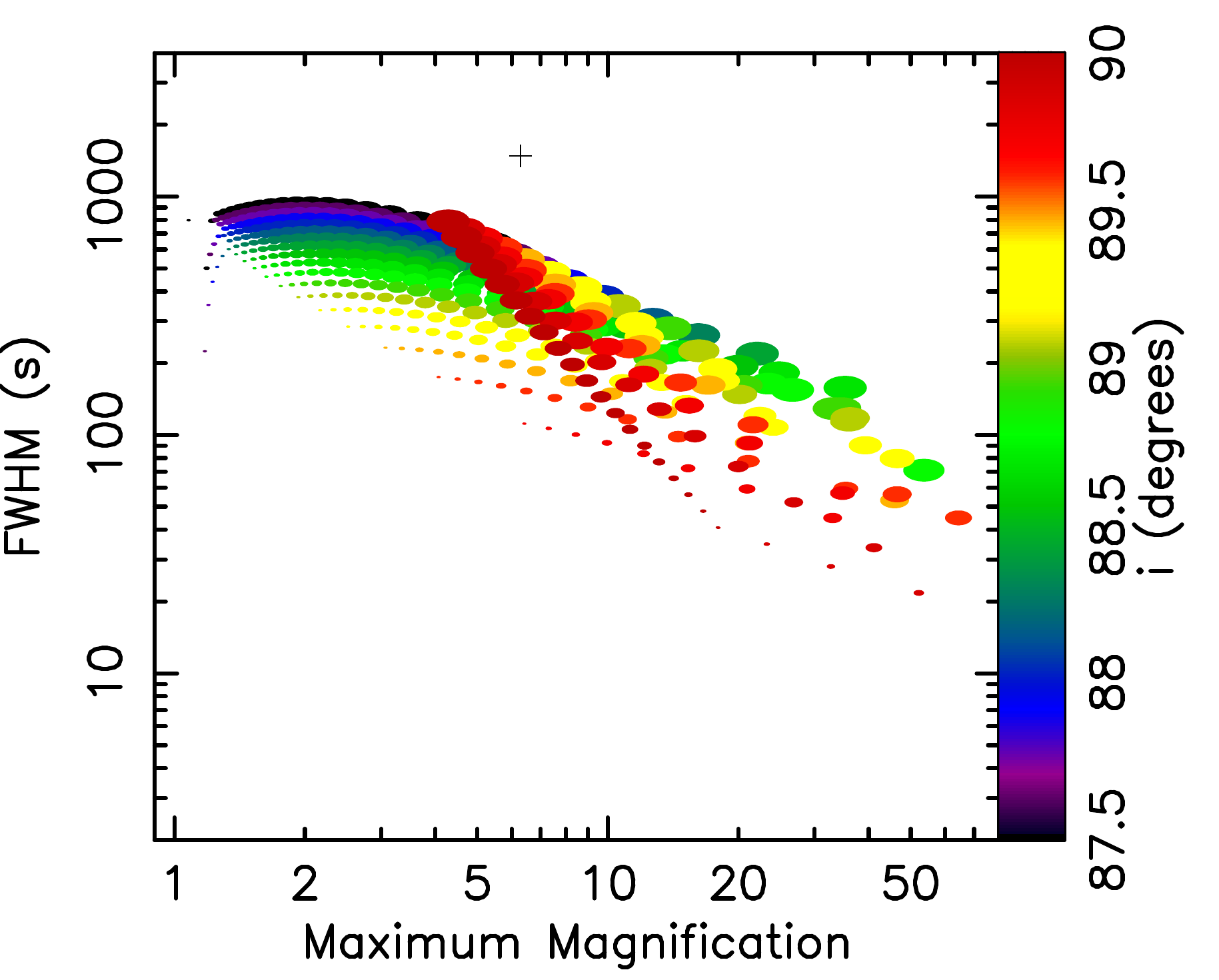}
\caption{Lensing flare FWHM versus peak magnification for two configurations in which we view the source and lens discs edge-on ($\gamma=90^\circ$). The BH spin parameter is $a=0.998$. Left: $\delta=70^\circ$, Right: $\delta=90^\circ$. Binary mass is varied in the range $M=5\times 10^4 - 5\times 10^6~M_\odot$, with larger markers corresponding to larger mass. Inclination angle is denoted by the colour code. The black cross corresponds to the QPEs in GSN 069.}
\label{fig:g90}
\end{figure*}

We additionally trial a number of parameter combinations with non-zero values of the misalignment angle, $\delta$, all assuming maximal spin. In order to maximise the flare amplitude, we trial parameter combinations in which the very inner (i.e. brightest) regions of the background mini-disc can be lensed without being occulted by the foreground mini-disc. In Fig. \ref{fig:g180}, we explore configurations whereby we view the underside of the foreground and background mini-discs (i.e. $\gamma=180$). The left and right panels are for $\delta=0.5^\circ$ and $\delta=1^\circ$ respectively. We see the same problem, in that the model cannot produce flares that are both large and long, only one or the other. In Fig. \ref{fig:g90}, we explore two configurations whereby we view both mini-discs exactly edge-on ($\gamma=90^\circ$). The left and right panels are for $\delta=70^\circ$ and $\delta=90^\circ$ respectively (plus we also tried $\delta=30^\circ$ with similar results).

\begin{figure}
\centering
\includegraphics[width=\columnwidth,trim=1.5cm 2.0cm 1.5cm 10.5cm,clip=true]{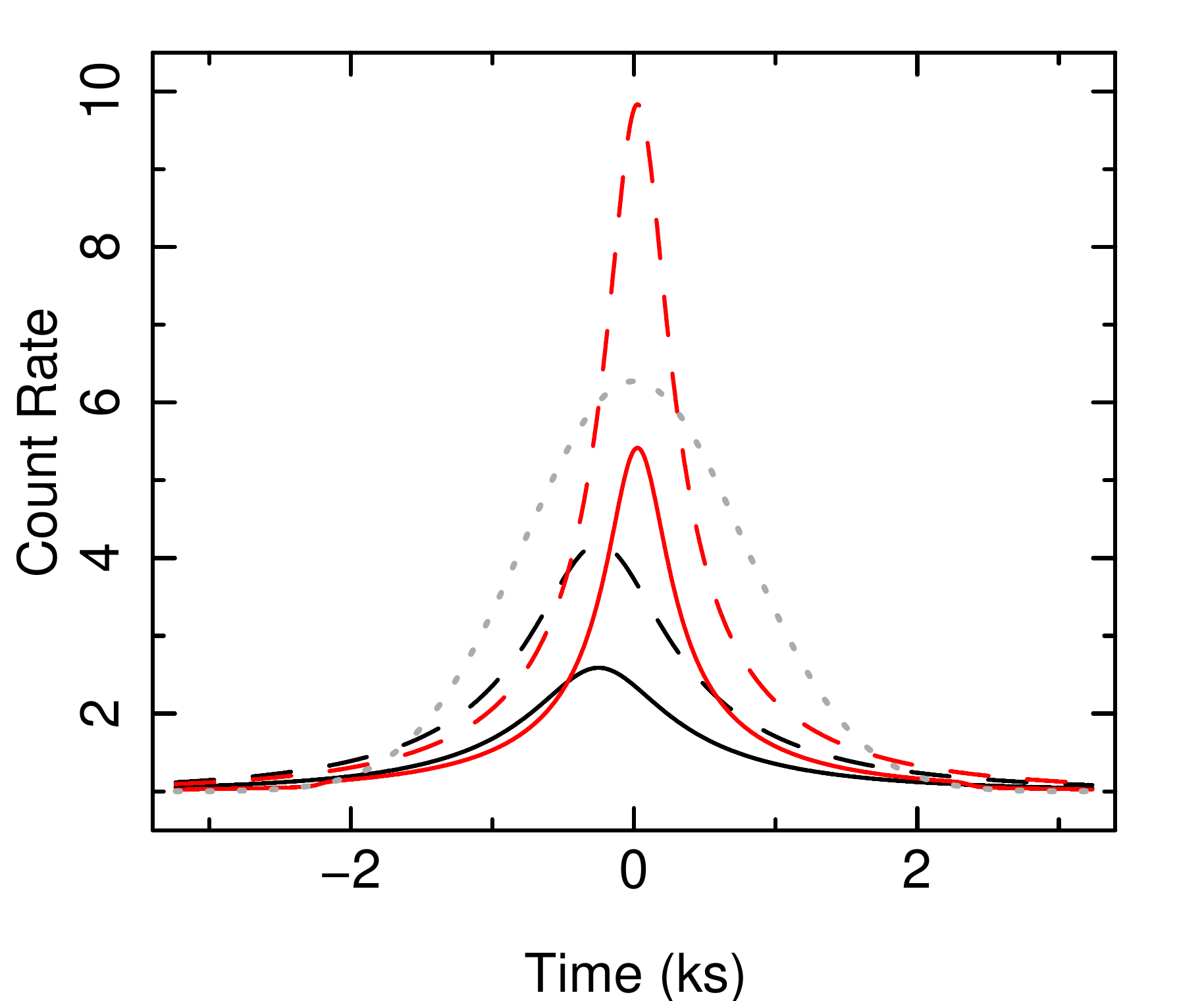}
\caption{$0.4-2$ keV light curves for the model parameters closest to reproducing the properties of the GSN 069 QPE flares. Black lines are the best aligned ($\delta=0$) model and red lines are the best overall model ($\delta=\gamma=90^\circ$). See text for the other parameters. Dashed lines are without occultation by the foreground mini-disc, and solid lines are with occultation. The grey dotted line represents the average GSN 069 flare profile.}
\label{fig:best}
\end{figure}

Of all the parameter combinations trialled, we have been unable to find one which produces a flare with amplitude $\mathcal{M} \ge 6$ and FWHM$\ge 1.5$ ks. In order to determine which parameter combination gets the closest to these values, we evaluate the statistic $\Delta = \sqrt{ (\mathcal{M}/6.3-1)^2 + (\mathrm{FWHM}/1.5~\mathrm{ks}-1)^2  }$ and take the combination with the minimum $\Delta$ to be the best. For the aligned model ($\delta=0$), $\Delta$ is minimised for $a=0.998$, $M=5\times 10^6~M_\odot$ and $i=87.63^\circ$ ($\Delta=0.6)$. The overall minimum $\Delta$ is achieved for $\delta=90^\circ$, $\gamma=90^\circ$, $a=0.998$, $M=5\times 10^6~M_\odot$ and  $i=87.5^\circ$ ($\Delta=0.59$). We plot the $0.4-2$ keV flare profile for both of these models in Fig. \ref{fig:best}. The black lines represent the aligned model, and the red lines the $\delta=\gamma=90^\circ$ model. Dashed lines are without occultation and solid lines are with it. We see that, at least in the aligned case, the model gets a lot closer to reproducing the GSN 069 QPE flares (represented by the grey dashed line) if occultation is ignored than when it is accounted for. It may therefore be possible for the model to work if the mini-disc were optically thin (i.e. more like an X-ray corona), which would give flare profiles somewhere between the dashed and solid lines. However, the observed spectrum is well described by a multi-temperature blackbody -- which requires the discs to be optically thick. Moreover, the energy dependence is still too subtle to explain the observations. We also note that the high mass employed for this plot implies that the binary in GSN 069 would have already merged by now ($t_\mathrm{merge}\approx 1.2$ yrs).

\section{Discussion}
\label{sec:discussion}

We have explored a self-lensing binary BH model for the QPEs observed from two low mass AGN: GSN 069 and RX J1301. We find that the $\sim 19$ yr evolution of the waiting time between consecutive QPEs observed for RX J1301 is consistent with the binary BH hypothesis as long as the current binary eccentricity is $\epsilon \gtrsim 0.16$. Since GW emission erodes eccentricity, lower eccentricity solutions are more likely, favouring lower binary masses; $M \lesssim 10^6~M_\odot$. The simulations of \citet[][see their Fig. 4]{Zrake2020} imply that an even lower mass of $M \lesssim 10^4~M_\odot$ is required for a $9$ hr period, $q\sim 1$ binary to have an eccentricity of $\epsilon \gtrsim 0.1$. We cannot place eccentricity limits on GSN 069, but the $\sim$equal spacing between QPEs favours $\epsilon \sim 0$.
In addition to orbital decay via GW emission reducing the orbital period over time, periastron precession will evolve the relative waiting times between the large and small flares in the two-flare QPE cycle, if $\epsilon > 0$. The model therefore predicts such an effect to occur for RX J1301 (periastron precession period of $\sim 1$ yr) but not for GSN 069. However, we have not been able to reproduce the flare profiles with the self-lensing model. Self-lensing flares are either the correct amplitude but too sharp, or the correct duration but too weak. Moreover, the model can reproduce stronger flares in harder energy bands due to the brighter parts of the background accretion disc radiating a harder spectrum, but the energy dependence of the predicted self-lensing flares is far more subtle than that observed for QPEs.

The discrepancy between the predicted and observed flare profiles is relatively small -- there are solutions with the correct amplitude but FWHM $\sim 2/5$ of the observed value, and those with the correct FWHM and amplitudes $\sim 2/5$ of the observed value. It is therefore possible that improving on some of the simplified assumptions in the model may in future enable it to reproduce the observed flare profiles. For example, we account for lensing by the foreground BH using the formula relevant to e.g. lensing of background galaxies by a foreground cluster \citep[e.g.][]{Walsh1979} or micro-lensing via chance alignments in our galaxy \citep[e.g.][]{Paczynski1986,Agol2002}, which ignores strong field effects \citep{Schnittman2018}. In particular, rays passing close to a spinning BH are preferentially concentrated to one side of the Einstein ring instead of being uniformly concentrated around it \citep[e.g.][]{ynogk}, which will likely increase the magnification and also the energy dependence of the flares, as different regions of the background mini-disc will be magnified more as the foreground BH passes in front of them. We also employ an approximate prescription for Shapiro delays. Whilst we do limit our parameter exploration to equal mass ratio binaries, we note that any relaxation of this assumption that has a large effect on the flare amplitude will also predict a bigger ratio between the amplitude of large and small flares than is observed. There are, however, many simplifying assumptions we have made about the properties of the mini-discs that could be relaxed in future. Allowing them to be optically thin would automatically increase the flare amplitudes by reducing occultation of the background disc by the foreground disc. However, this would need to be justified by demonstrating that the spectrum of the QPE sources is consistent with being non-thermal. The discs may also have a non-zero scale-height, may not be perfectly aligned with their BH spin axis, may be mildly truncated outside of the ISCO, may have a shallower emissivity profile than the canonical $r^{-3}$ relation, their outer radii may be different from what we assume here, or the gas streams between mini-discs and from circumbinary disc to mini-discs that we ignore here may in reality be important. We also do not include a compact corona close to each BH in our model. This is driven by the X-ray spectrum of GSN 069 and RX J1301 being disc-dominated, although there is evidence for a weak power-law component in the spectrum of RX J1301 \citep{Middleton2015}. It is therefore still possible for the self-lensing model to yet reproduce the observed QPE profiles, but it is harder to imagine the model being able to reproduce the flare profiles for a reasonably large part of parameter space so as not to be fine-tuned, or reproducing the strong energy dependence of the observed flares. We therefore conclude that the two QPE sources could in principle be binary BH systems, but that the mechanism for the flares is unlikely to be lensing of the background mini-disc by the foreground BH.

It is also important to note that very close binaries, within $\sim 100$ yrs of merger, must be rare in the local Universe, since this phase is a small fraction of the overall binary lifetime. This favours lower binary mass in our model, since lower mass binaries with a given period are further from merger (see e.g. Fig. \ref{fig:tmerge}). This requirement is in tension with the flare profile modelling, which is closer to reproducing the observed flares for high lens mass. In any case, this consideration means that the model requires QPEs with short flare spacings to be rare. Similar sources with much larger spacings between flares may be much more common, but such flares are potentially easier to miss.

The discovery of close binary massive BH systems in the GW inspiral phase is a very important goal in astrophysics that as yet remains elusive. Such systems are thought to be important diagnostics of galaxy evolution, since galaxy mergers result in the central BHs of the two constituent galaxies forming an initially wide binary that eventually decays down to $\sim 1$ pc separation via dynamical friction \citep[e.g.][]{Barnes1996,Barnes2002,Mayer2013}. It is unclear how binary systems then become close enough for GW emission to efficiently evolve the orbit (\textit{the final parsec problem}: \citealt{Begelman1980,Milosavljevic2003,Armitage2005}), eventually resulting in a merger with a detectable GW signal. Discovery of a close binary system via electromagnetic emission will therefore definitively confirm that at least nature knows how to solve the final parsec problem, and will provide important constraints on the expected merger rates for \textit{LISA}.

Wide (separation $\gg 1$ pc) accreting BH-BH binary systems can be identified using high resolution observations
(X-ray: \citealt{Komossa2003,Comerford2015}; optical: \citealt{Comerford2015,Goulding2019,Kollatschny2020}; radio: \citealt{Rodriguez2006a,Fu2015,Kharb2017}), but other indirect methods are required to identify close binaries in the GW inspiral phase (separation $\lesssim 0.01$ pc). Proposed diagnostics include a broad dip in the disc blackbody spectrum caused by the cavity opened up by tidal forces (\citealt{Roedig2014} -- although this dip was not recovered in the simulations of \citealt{Farris2015} and \citealt{d'Ascoli2018} since emission from the accretion streams compensates for the presence of the cavity), a `ripple shaped' iron K$\alpha$ line caused by Doppler shifts from the two orbiting mini-discs \citep{McKernan2013}, a periodicity due to Doppler boosting of the orbiting mini-discs \citep{D'Orazio2015} and quasi-periodic oscillations due to interactions between the mini-discs and the circumbinary disc \citep{MacFadyen2008,D'Orazio2013,Farris2014,Farris2015,Shi2015}. More than $100$ candidate sub-pc binaries have been identified by searching for periodicities in large optical surveys (e.g. \citealt{Lehto1996,Graham2015,Graham2015a,Charisi2016}; see Fig. 2 in \citealt{D'Orazio2018} for a compilation). However, none have been definitively confirmed as binary systems, largely because AGN ubiquitously display strong red noise variability that is very likely to appear quasi periodic for several cycles, particular when large data sets are considered without adequately accounting for the large number of trials \citep{Vaughan2016}. Also, true quasi-periodic behaviour is not exclusive to binary BH-BH systems \citep[e.g.][]{Ingram2019b}. Indeed, a sample of 7 objects identified as binary candidates via periodicity claims were recently found to have X-ray spectral properties indistinguishable from the general AGN population \citep{Saade2020}. Self-lensing flares have been suggested as a distinctive signature of high inclination accreting binary BH systems \citep{Haiman2017,D'Orazio2018}, and one candidate system, nicknamed Spikey, has been identified in archival \textit{Kepler} data \citep{Hu2020,Kun2020}. The light curve of Spikey can be modelled with a self-lensing flare plus Doppler modulation from a binary system with parameters $\epsilon \approx 0.5$, $P\approx 418$ d, $M\approx 3\times 10^7~M_\odot$, $i\approx 82^\circ$, and the model predicts another lensing flare to happen this year.

The modelling we have presented here builds upon earlier work \citep{Haiman2017,D'Orazio2018} by including more relativistic effects and, most crucially, including occultation by the foreground mini-disc. This gives rise to a variety of behaviours: different parameters lead to flares (e.g. Fig. \ref{fig:lc_a0_d0}), eclipses (e.g. Fig. \ref{fig:lc_a0_d30}), and eclipses with a central narrow flare (e.g. Fig. \ref{fig:lc_a0_d10}). In particular, the eclipse with a central flare is an extremely distinctive feature that can be searched for in X-ray survey data (e.g. \textit{XMM-Newton}, \textit{eROSITA}). In future, we will extend our modelling to determine if any similarly distinctive features are predicted in the optical, which can then be searched for in surveys such as \textit{ZTF} and \textit{LSST}. For example, two very sharp flares may occur during the eclipse as the outer part of the background mini-disc twice passes close to the Einstein radius of the foreground BH (i.e. first the right hand side of the disc as we see it, then the left hand side), or lensing of emission from the circumbinary disc by the two BHs could cause a flare if the binary plane is misaligned with the circumbinary disc. If such a feature were to be discovered, it would constitute very strong evidence for the presence of a binary SMBH system. We note that such a discovery would require the molecular torus associated with the AGN to either be misaligned with the binary plane (which is plausible, for instance the AGN accretion disc orientation does not appear to be correlated with that of the host galaxy's stellar disc: \citealt{Middleton2016}), or to have not yet fully formed (under a paradigm whereby most AGN activity is triggered by galaxy mergers).

If the QPEs from GSN 069 and RX J1301 are not due to self-lensing, then what are they caused by? It could be that the system is a binary, but with a very low mass ratio $q \lesssim 10^{-2}$ such that the flares are instead caused by the secondary crashing through the primary's accretion disc twice per orbital period -- as is suggested for the much more massive ($M \sim 1.71 \times 10^{10}~M_\odot$, $q\sim 6\times 10^{-3}$) AGN OJ287 \citep{Lehto1996}. We note that in this scenario, the requirement for QPEs to be very rare is relaxed a lot compared with the lensing model since such a system would be far from merging and does not need to be viewed from a special angle, plus minor mergers are more frequent than major mergers. Alternatively, \citet{King2020} suggested that the QPEs in GSN 069 result from a `near miss' tidal disruption event (TDE) of a red giant whose envelope was stripped, leaving behind a white dwarf core (current mass $\sim 0.21~M_\odot$) on an eccentric orbit (current eccentricity $\epsilon_0\sim 0.94$) around a single $M\sim 4\times 10^5~M_\odot$ BH. In this case, the QPE flares occur due to enhanced accretion each time the orbit reaches pericenter. This model nicely explains the factor $\gtrsim 240$ rise in the quiescent flux level of GSN 069 between a \textit{ROSAT} non-detection in 1994 and the first \textit{XMM-Newton} detection in 2010 (i.e. the near-miss TDE s assumed to have occurred at some time during the intervening 16 years), followed by the subsequent onset of QPEs and slow decay of the quiescent flux level. In our self-lensing picture, there is no ready interpretation for the rise in the quiescent flux level, but the sudden onset of QPEs could in principle be explained by the orbital separation becoming small enough for lensing flares to begin (the lensing amplitude can be a steep function of impact parameter for certain values of impact parameter). The failed TDE model, however, struggles to explain the symmetric nature of the QPEs (accretion flares have a fast rise, exponential decay profile). It also cannot reproduce the long-waiting-time-followed-by-short-waiting-time behaviour of the QPEs, particularly for RX J1301 in which this behaviour is much more pronounced.

It has alternatively been suggested \citep{Miniutti2019} that QPEs could be due to limit-cycle accretion rate variations, analogous to the `heartbeat' oscillations observed from Galactic X-ray binaries GRS 1915+105 \citep{Belloni2000},  IGR J17091--3624 \citep{Altamirano2011} and the Rapid Burster \citep{Bagnoli2015}, and the ultra-luminous X-ray source NGC 3621 \citep{Motta2020}. However, the duration of QPEs are far fewer dynamical time scales than the duration of any heartbeat flares in lower mass systems, and it is difficult to imagine how accretion variation could lead to such sharp, symmetric flares on a timescale $\sim$one hundredth of a viscous timescale. \citet{Miniutti2019} argue that `changing-look' AGN also vary faster than the viscous timescale and therefore standard accretion theory may simply be missing something, but we do not know that changing look AGN are driven by intrinsic accretion changes. Another model, suggested for RX J1301 before the GSN 069 QPEs were discovered, is the `Swiss cheese' model of \citet{Middleton2015}. This assumes that a low mass AGN is surrounded by an optically thick shroud of orbiting material. A hole in the shroud then reveals the centre of the disc each time it orbits into the line of sight. This model can qualitatively describe the spectral evolution of the only QPE that was known at the time (the single full flare in the 2000 \textit{XMM-Newton} observation of RX J1301). However, it does not explain the long-waiting-time-followed-by-short-waiting-time behaviour, and it also seems unlikely that a hole in a similar part of the shroud would still be there 19 years after the first QPE was observed.

\section{Conclusions}
\label{sec:conclusions}

We have explored a self-lensing binary SMBH model for the sharp flares (QPEs) exhibited by the low mass AGN GSN 069 and RX J1301. Our model can reproduce the observed QPE recurrence times if the current eccentricity of RX J1301 is $\epsilon_0 \gtrsim 0.16$, and the evolution of recurrence times in observations of RX J1301 separated by $\sim 19$ years suggests a total binary mass $\lesssim 10^6~M_\odot$ that will merge within the next $\sim 1000$ yrs. However, we have been unable to reproduce the observed flare profiles with our gravitational self-lensing model -- predicting either flares of the correct amplitude that are too sharp, or flares of the correct duration that are too low in amplitude. We conclude that self-lensing is unlikely to be the mechanism behind the observed QPEs, although we note that improving upon some of the simplified assumptions employed in this paper may yet enable to model to reproduce the observed flares. Several models have been suggested for QPEs, but none can currently reproduce the full array of observational properties.

Self-lensing remains a promising means for discovering binary SMBH systems in future. Our modelling extends upon previous work by including more GR effects and, most significantly, occultation of the background mini-disc by the foreground mini-disc. We find three distinctive behaviours: i) lensing flares with occultation slightly reducing the peak magnification, ii) partial eclipses (in-eclipse flux originates from the foreground mini-disc) with no associated lensing flare, and iii) partial eclipses with a very sharp in-eclipse lensing flare. All three behaviours are distinctive diagnostics of binary SMBH systems that can be searched for in current and future surveys (e.g. \textit{eROSITA}, \textit{ZTF}, \textit{LSST}) to identify binaries and to place strong constraints on their properties.

\section*{Acknowledgements}

We thank the anonymous referee for insightful comments that improved the paper.
We acknowledge valuable discussions with Giovanni Miniutti and Margherita Giustini. A.I. acknowledges support from the Royal Society and fruitful discussions with Andrew King, Matt Middleton, Philipp Podsiadlowski, Andy Mummery, and Rob Fender. SA acknowledges support from the STFC (grant numbers ST/N000919/1 and ST/R004846/1).

\section*{Data availability}

The data underlying this article, including all source code, will be shared on reasonable request to the corresponding author.




\bibliographystyle{mnras}
\bibliography{biblio} 




\appendix

\section{Details of lensing flares model}

\subsection{Coordinate system}
\label{sec:coord}

The observer's line of sight is
\begin{equation}
    \mathbf{\hat{o}}= \sin i \cos\phi_i ~\mathbf{\hat{x}_{\rm bin}} + \sin i \sin\phi_i ~\mathbf{\hat{y}_{\rm bin}} + \cos i  ~\mathbf{\hat{z}_{\rm bin}},
\end{equation}
such that projected distance on the observer's sky between the two BHs is smallest when the orbital phase is $\phi_*=\phi_i$. The rotation axis of the background (source) mini-disc is
\begin{equation}
    \mathbf{\hat{z}_s}= \sin \delta_s \cos(\phi_i-\gamma_s) ~\mathbf{\hat{x}_{\rm bin}} + \sin \delta_s \sin(\phi_i-\gamma_s) ~\mathbf{\hat{y}_{\rm bin}} + \cos \delta_s  ~\mathbf{\hat{z}_{\rm bin}}.
\end{equation}
The expression for $\mathbf{\hat{z}_\ell}$ is the same with $s$ subscripts replaced by $\ell$ subscripts.

The image plane is perpendicular to $\mathbf{\hat{o}}$. It is convenient to define the vertical axis of the image plane as the projection of the BH spin axis on the image plane. Since the two BHs can have different spin axes to one another, we define one vertical axis for the background BH, $\mathbf{\hat{\beta}_s}$, and another for the foreground BH, $\mathbf{\hat{\beta}_\ell}$. The horizontal axis of the image plane for the background BH is therefore
\begin{equation}
    \mathbf{\hat{\alpha}_s} = \frac{\mathbf{\hat{z}_s}  \times \mathbf{\hat{o}} } {|\mathbf{\hat{z}_s} \times \mathbf{\hat{o}}|},
\end{equation}
and the vertical axis is $\mathbf{\hat{\beta}_s} = \mathbf{\hat{o}} \times \mathbf{\hat{\alpha}_s}$; with similar expressions for $\mathbf{\hat{\alpha}_\ell}$ and $\mathbf{\hat{\beta}_\ell}$. In the main text, we simply use $\alpha$, $\beta$ and $\mathbf{b}$ and specify which definition of the image plane we are referring to.

The vector $\mathbf{b_*} = \alpha_* \mathbf{\hat{\alpha}_s} + \beta_* \mathbf{\hat{\beta}_s}$, where $\alpha_* = \mathbf{r_*} \cdot \mathbf{\hat{\alpha}_s}$ and $\beta_* = \mathbf{r_*} \cdot \mathbf{\hat{\beta}_s}$, points on the image plane from the centre of the source BH to the centre of the lens BH. The vectors $\mathbf{b_s}$ and $\mathbf{b_\ell}$ are also defined on the image plane. $\mathbf{b_s}$ points from the source BH to a given point on the source disc, 
and $\mathbf{b_\ell}$ points from the lens BH to 
a given point on the lens disc. The vector that 
points from the lens BH to a given point on the 
source disc is therefore 
$\mathbf{b_s}-\mathbf{b_*}$. We can determine if 
a given point on the background (source) is 
eclipsed by the foreground (lens) disc by 
comparing the vectors $\mathbf{b_\ell}$ and 
$\mathbf{b_s}-\mathbf{b_*}$.

\subsection{Shapiro Delays}
\label{sec:shap}

\begin{figure}
\centering
\includegraphics[width=\columnwidth,trim=8.0cm 1.5cm 1.0cm 1.0cm,clip=true]{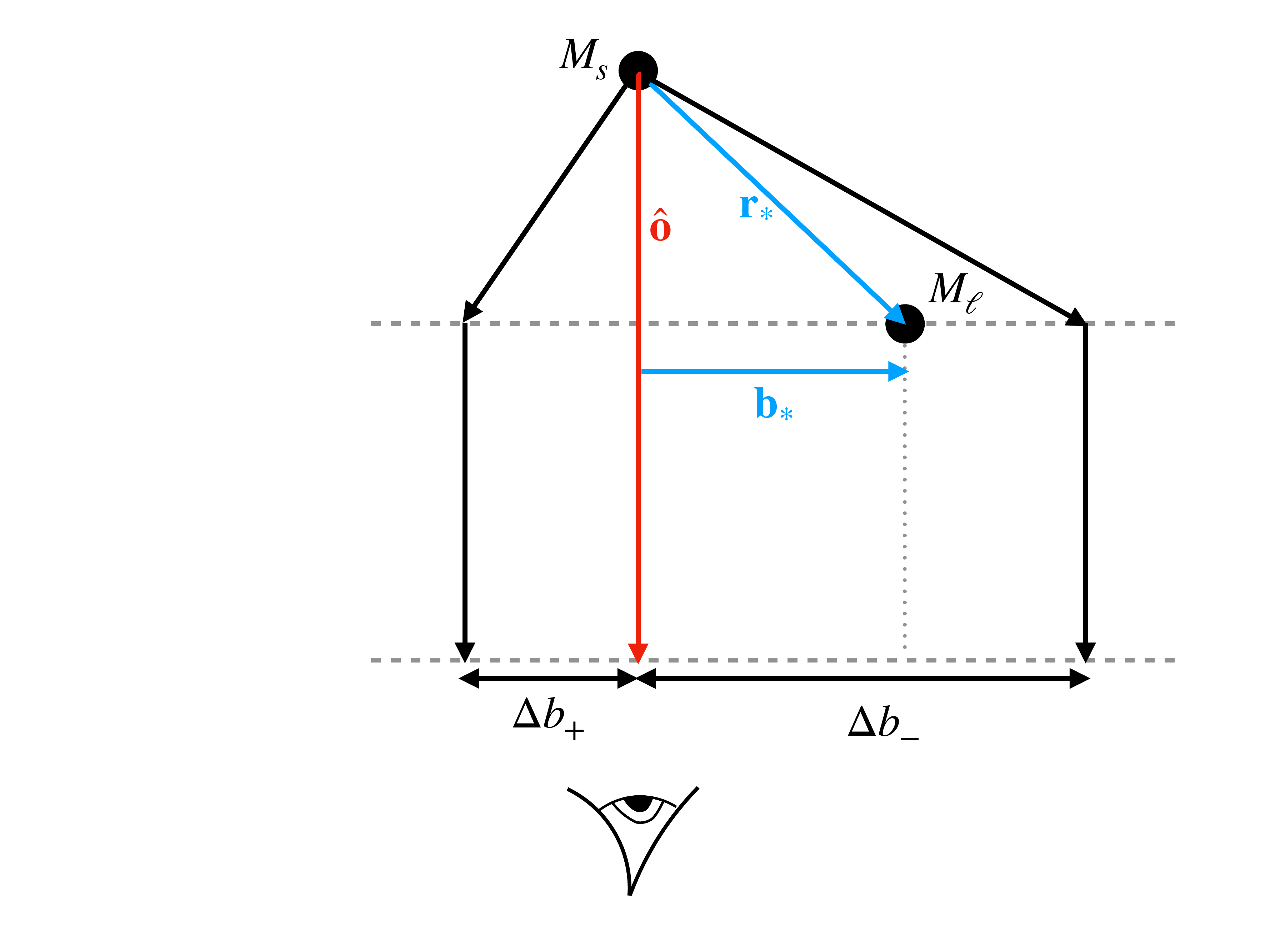}
\caption{Schematic illustrating the path length of photons for the internal image (black arrows on the right) and the external image (black arrows on the left). See text for further details.}
\label{fig:shap}
\end{figure}

Fig. \ref{fig:shap} illustrates the approximate path of rays from a source very close to the background BH (labelled $M_s$), around the foreground BH (labelled $M_\ell$), and finally towards the observer (bottom). The vector $\mathbf{r_*}$ points from the background BH to the foreground BH, $\mathbf{b_*}$ is the projection of $\mathbf{r_*}$ on the image plane, and $\mathbf{\hat{o}}$ points from the source BH towards the observer. In the absence of the foreground BH, the image would be at the origin in this example. Instead, there are two images: the internal image at $\Delta b_-$, and the external image at $\Delta b_+$. The rays for each image, denoted by the black arrows, have a longer path length than an un-lensed ray, that would travel along $\mathbf{\hat{o}}$ (red arrow). This extra path length is
\begin{equation}
    c \Delta t_{\pm}(\phi_*) = \sqrt{ (\mathbf{r_*}(\phi_*)\cdot \mathbf{\hat{o}})^2 + (\Delta b_\pm)^2} - \mathbf{r_*}(\phi_*)\cdot \mathbf{\hat{o}},
\end{equation}
where $\Delta t_\pm$ is the Shapiro delay. This is an approximate treatment of the Shapiro delay since in reality the geodesics are not straight lines with a sharp bend, and there are also Shapiro delays due to lensing around the background BH. Also, the above formula assumes that the source is very small compared with the binary separation. In other words, it assumes that $r \ll r_*$ for all disc patches. This is not the case for the outer disc ($r_{\rm out}=0.25~r_*$), but the X-ray flux is dominated by light from the inner disc due to the steep gravitational emissivity. A more accurate treatment of Shapiro delays would therefore be required for the optical signal, for which the emissivity is less centrally peaked.

We calculate $\Delta t_\pm(\phi_*)$ for the orbital phase corresponding to the time the photon was emitted, $\phi_*(t_{\rm em})$. The photon crosses the top grey dashed line in Fig. \ref{fig:shap} (after which all geodesics have the same light travel time) at time $t=t_{\rm em} + \Delta t_\pm[\phi_*(t_{\rm em})]$. Practically, we start by specifying the time $t$ (which is the same for every pixel of the image), and use an iteration scheme to calculate $t_{\rm em}$
\begin{equation}
    t_{{\rm em}(i+1)} = t - \Delta t_\pm [\phi_*(t_{{\rm em}(i)}],
\end{equation}
where we set $t_{{\rm em}(0)}=t$. The iteration scheme converges very quickly, typically after a few steps.

\subsection{Disc Spectrum}
\label{sec:spec}

For each time bin and for each patch of the background mini-disc, we calculate the position of the two images of the patch on the image plane, and assess if either of these are eclipsed by the foreground disc. The specific flux of the background mini-disc at time $t$ is
\begin{equation}
    F_s(E,t) \propto \int_{\alpha_s,\beta_s} \left[\eta_+(t) \mathcal{M}_+(t) + \eta_-(t) \mathcal{M}_-(t) \right] g^3  I(E/g,T) d\alpha_s d\beta_s,
\end{equation}
where $\eta_{\pm}$ is $0$ if the image of the background disc patch is eclipsed by the foreground disc and $1$ otherwise. $\mathcal{M}_\pm$ is the magnification, given by Equation (\ref{eqn:mag}) in the main text. $I(E,T)$ is the specific intensity radiated by the disc at coordinates $r$, $\phi$, which we assume to be a blackbody with the temperature profile given in the main text. $g$ is the blueshift experienced by a photon travelling from the disc patch to the observer (see Equation 4 of \citealt{Ingram2017}).


\bsp	
\label{lastpage}
\end{document}